\documentclass[11pt,a4paper]{article}

\usepackage{amssymb}

\usepackage[dvips]{graphicx}

\begin{document}

\title{\bf NSVZ scheme with the higher derivative regularization
for ${\cal N}=1$ SQED}

\author{
A.L.Kataev\\
{\small{\em Institute for Nuclear Research of the Russian Academy of Science,}}\\
{\small {\em 117312, Moscow, Russia}},\\
\\
K.V.Stepanyantz\\
{\small{\em Moscow State University}}, {\small{\em  Physical
Faculty, Department  of Theoretical Physics}}\\
{\small{\em 119991, Moscow, Russia}}}

\maketitle

\begin{abstract}
The exact NSVZ  relation between a $\beta$-function of ${\cal
N}=1$ SQED and an anomalous dimension of the matter superfields is
studied  within the Slavnov higher derivative regularization
approach. It is shown that if the renormalization group functions
are defined in terms of the bare coupling constant, this relation
is always valid. In  the renormalized theory the
NSVZ relation is obtained in the momentum subtraction scheme
supplemented by a  special finite renormalization. Unlike the
dimensional reduction, the higher derivative regularization
allows to fix this finite renormalization. This is made by
imposing the conditions $Z_3(\alpha,\mu=\Lambda)=1$ and
$Z(\alpha,\mu=\Lambda)=1$ on the renormalization constants of
${\cal N}=1$ SQED, where $\Lambda$ is a parameter in the higher
derivative term. The results are verified by the  explicit
three-loop calculation. In this  approximation we relate the
$\overline{\mbox{DR}}$ scheme and the NSVZ scheme defined within
the higher derivative approach by the finite renormalization.

\end{abstract}

\unitlength=1cm

Keywords: higher covariant derivative regularization,
supersymmetry, $\beta$-function, subtraction scheme.

%%%%%%%%%%%%%%%%%%%%%%%%%%%%%%%%%%%

\section{Introduction}
\hspace{\parindent}

Supersymmetry is a beautiful concept of quantum field theory
\cite{Ogievetsky:1975nu}, which is studied both by theoreticians
and by experimentalists. Investigations of SUSY models reveal
interesting  theoretical features (see
e.g.\cite{West:1990tg,Buchbinder:1998qv,Diakonov:2010zzb}). Among
them is the finiteness of the   ${\cal N}=4$ supersymmetric Yang
Mills (SYM) theory. As the consequence, this theory obeys the
conformal symmetry even at the quantum level due to  vanishing of
the renormalization group (RG) $\beta$-function. This was proved
in all orders of the perturbation theory in
\cite{Grisaru:1982zh,Mandelstam:1982cb,Brink:1982pd,Howe:1983sr}
after the explicit analytical  three-loop calculation of Ref.
\cite{Avdeev:1980bh}, made with the  dimensional reduction (DRED)
regularization \cite{Siegel:1979wq} (this  calculation was also
confirmed in \cite{Grisaru:1980nk,Caswell:1980ru}). DRED is a
modification of the dimensional regularization method
\cite{'tHooft:1972fi,Bollini:1972ui,Ashmore:1972uj,Cicuta:1972jf}.
It is applied in SUSY theories since the dimensional
regularization explicitly breaks SUSY \cite{Delbourgo:1974az}.
Renormalization of SUSY theories is usually made using  a variant
of the modified minimal subtraction scheme $\overline{\mbox{MS}}$
\cite{Bardeen:1978yd}\footnote{The  detailed definition  of the
$\overline{\mbox{MS}}$ scheme  is given  in
Ref.\cite{Kataev:1988sq}.}, namely in  the $\overline{\mbox{DR}}$
scheme. However, it is well known that DRED is not mathematically
consistent \cite{Siegel:1980qs}. Its inconsistency leads to the
loss of explicit SUSY \cite{Avdeev:1981vf}, which will  be broken
by quantum corrections in higher loops
\cite{Avdeev:1982np,Avdeev:1982xy}. Nevertheless, a recent
analytical  calculation \cite{Velizhanin:2010vw} demonstrates that
the DRED regularization and the $\overline{\mbox{DR}}$ scheme are
working in the ${\cal N}=4$ SYM theory up to the four-loop
approximation and give the  vanishing result for its
$\beta$-function.

SUSY also leads  to the absence of radiative corrections  to the
$\beta$-function of  ${\cal N}=2$ SYM theories starting from the
two-loop approximation \cite{Howe:1983wj}. In this case vanishing
of the two- and three-loop contributions to the $\beta$-function
was explicitly demonstrated in \cite{Avdeev:1981ew} using the
$\overline{\mbox{DR}}$ scheme. However, the calculations made in
Ref. \cite{Avdeev:1982np}, even after correcting them in Refs.
\cite{Jack:2007ni,Velizhanin:2008rw}, show that the
$\beta$-function obtained from the fermion-fermion-vector,
scalar-scalar-vector and ghost-ghost-vector vertexes vanishes, but
obtaining the $\beta$-function from the fermion-fermion-scalar
vertex gives a non-vanishing result in the three-loop
approximation \cite{Velizhanin:2008rw}.  This result reveals
unsolved theoretical problems which appear if DRED is used for the
regularization of SUSY theories.

${\cal N}=1$ SUSY leads to the absence of divergent quantum
corrections to the superpotential \cite{Grisaru:1979wc}. Another
interesting feature of ${\cal N}=1$ SYM models is  existence of
the exact relation between the $\beta$-function and the anomalous
dimension of the matter superfields, derived in Refs.
\cite{Novikov:1983uc,Novikov:1985rd,Shifman:1986zi}. This relation
is usually called "the exact Novikov, Shifman, Vainshtein, and
Zakharov (NSVZ) $\beta$-function". It was first obtained in terms
of the  non-renormalized bare coupling constant $\alpha_0$ for SYM
models in Refs. \cite{Novikov:1983uc,Novikov:1985rd} and for the
${\cal N}=1$ supersymmetric electrodynamics (SQED) in Refs.
\cite{Vainshtein:1986ja,Shifman:1985fi}. For ${\cal N}=1$ SQED the
NSVZ $\beta$-function is

\begin{equation}\label{NSVZ}
\beta(\alpha_0) =
\frac{\alpha_0^2}{\pi}\Big(1-\gamma(\alpha_0)\Big).
\end{equation}

However, a general  problem  arises, whether it is possible to
specify a concrete subtraction scheme, which leads to the exact
NSVZ $\beta$-function, if the RG functions are defined through the
renormalized coupling constant. With DRED, supplemented by the
$\overline{\mbox{DR}}$ subtraction scheme, this problem was
studied in Refs.
\cite{Jack:1996vg,Jack:1996cn,Jack:1998uj,Jack:2007ni}. The
results of the one- \cite{Ferrara:1974pu} and two-loop
\cite{Jones:1974pg} calculations agree with the NSVZ
$\beta$-function, because a two-loop $\beta$-function and a
one-loop anomalous dimension are scheme independent in theories
with a single coupling constant. In higher orders
\cite{Jack:1996cn,Jack:2007ni,Harlander:2006xq} the exact NSVZ
$\beta$-function for the RG functions defined in terms of the
renormalized coupling constant can be obtained with the
$\overline{\mbox{DR}}$ scheme after an additional finite
renormalization. This finite renormalization should be fixed in
each order of the perturbation theory, starting from the
three-loop approximation
\cite{Jack:1996vg,Jack:1996cn,Jack:1998uj,Jack:2007ni}. However,
there is no general prescription, how one should construct this
finite renormalization using the $\overline{\mbox{DR}}$ scheme.
Investigations of quantum corrections using other regularizations
\cite{Shifman:1985tj,Mas:2002xh} are usually made in one- and
two-loop approximations.

In this paper we study in details, how the exact relation
(\ref{NSVZ}) can be obtained using the Slavnov higher derivative
regularization \cite{Slavnov:1971aw,Slavnov:1972sq}, which is
mathematically consistent and does not break the supersymmetry in
all orders \cite{Krivoshchekov:1978xg,West:1985jx}. Therefore,
this regularization has theoretical advantages over the
mathematically inconsistent DRED. Application of the higher
derivative regularization to the evaluation of quantum corrections
in ${\cal N}=1$ SUSY theories reveals one more interesting
feature: integrals needed for obtaining a $\beta$-function defined
in terms of the bare coupling constant are integrals of total
derivatives
\cite{Soloshenko:2003nc,Pimenov:2009hv,Stepanyantz:2011zz,Stepanyantz:2011wq}
and even double total derivatives
\cite{Smilga:2004zr,Stepanyantz:2011bz,Stepanyantz:2012zz,Stepanyantz:2012us}.
This implies that one of the loop integrals can be calculated
analytically, and a $\beta$-function in a $L$-loop approximation
can be related with an anomalous dimension of the matter
superfields in the $(L-1)$-loop approximation
\cite{Stepanyantz:2011jy}. As a consequence, the NSVZ
$\beta$-function can be naturally obtained for the RG functions
defined in terms of the bare coupling constant $\alpha_0$ in the
case of using  the higher derivative  regularization. However, if
the RG functions are defined in terms of the renormalized coupling
constant, the NSVZ $\beta$-function is obtained only in a special
subtraction scheme, which is constructed in this paper for ${\cal
N}=1$ SQED regularized by higher derivatives. Unlike DRED, the
higher derivative regularization allows to construct the NSVZ
scheme by imposing the boundary conditions on the renormalization
constants. The results are verified by an explicit three-loop
calculation. In this approximation we also relate the
$\overline{\mbox{DR}}$ scheme with the NSVZ scheme obtained with
the higher derivatives regularization.

The paper is organized as follows: In Sect.
\ref{Section_Regularization} we introduce the higher derivative
regularization and remind how the NSVZ $\beta$-function can be
obtained using this regularization, if the RG functions are
defined in terms of the bare coupling constant and the
Pauli--Villars masses are proportional to the parameter $\Lambda$
in the higher derivative term. However, after a rescaling of the
Pauli--Villars masses which depends on the bare coupling constant
the NSVZ relation for these RG functions is no longer valid. This
is discussed in Sect. \ref{Section_Rescaling}. A standard
definition of the RG functions is recalled in Sect.
\ref{Section_Boundary_Conditions}. In general, these RG functions
(defined in terms of the renormalized coupling constant) do not
satisfy the NSVZ relation. However, there is an NSVZ scheme in
which the NSVZ relation is valid. In Sect.
\ref{Section_Boundary_Conditions} we demonstrate that this scheme
is obtained by imposing a special boundary conditions on the
renormalization constants, and the RG functions defined by two
different ways coincide in this scheme. Changing the boundary
conditions it is possible to change the relation between the
$\beta$-function and the anomalous dimension. All these results
are verified by an explicit three-loop calculation with the higher
derivative regularization in Sect.
\ref{Section_Explicit_Three_Loop}. In Sect.
\ref{Section_DRED_NSVZ} we compare the results obtained in the
$\overline{\mbox{DR}}$-scheme with the results obtained with the
higher derivative regularization in the NSVZ scheme. Some
technical details are discussed in the Appendexes.

\section{Derivation of the NSVZ $\beta$-function for ${\cal N}=1$ SQED}
\hspace{\parindent}\label{Section_Regularization}

In this section we recall how the NSVZ $\beta$-function can be
obtained with the higher derivative regularization for ${\cal
N}=1$ SQED. It is convenient to describe this theory in terms of
${\cal N}=1$ superfields \cite{West:1990tg,Buchbinder:1998qv}.
Then in the massless limit the action of ${\cal N}=1$ SQED is
written as

\begin{equation}\label{Action}
S = \frac{1}{4e_0^2}\mbox{Re}\int d^4x\,d^2\theta\,W^a W_a +
\frac{1}{4} \int d^4x\,d^4\theta\,\Big(\phi^* e^{2V}\phi +
\widetilde\phi^* e^{-2V} \widetilde\phi\Big),
\end{equation}

\noindent where $e_0$ is a bare coupling constant, $\phi$ and
$\widetilde\phi$ are (bare) chiral superfields, and $V$ is a real
gauge superfield.

In order to regularize this theory by higher derivatives, it is
necessary to insert into the first term of Eq. (\ref{Action}) a
regularizing function $R$, such that $R(0)=1$ and
$R(\infty)=\infty$, which contains higher derivatives
\cite{Slavnov:1971aw,Slavnov:1972sq}:

\begin{equation}
\frac{1}{4e_0^2}\mbox{Re}\int d^4x\,d^2\theta\,W^a W_a
\to\frac{1}{4e_0^2} \mbox{Re}\int d^4x\,d^2\theta W^a
R(\partial^2/\Lambda^2) W_a.
\end{equation}

\noindent For example, it is convenient to choose this function as
$R = 1+\partial^{2n}/\Lambda^{2n}$, where $\Lambda$ is a
dimensionful parameter.  Then the divergences remain only in the
one-loop approximation \cite{Faddeev:1980be}. According to the
standard prescription they should be regularized by inserting the
Pauli--Villars determinants into the generating functional
\cite{Slavnov:1977zf}:

\begin{equation}\label{Generating_Functional}
Z[J,j,\widetilde j] = \int DV\, D\phi\,
D\widetilde\phi\,\prod\limits_{I=1}^n
(\det(V,M_I))^{c_I}\exp\Big(i S_{\mbox{\scriptsize reg}} + i
S_{\mbox{\scriptsize gf}} + i S_{\mbox{\scriptsize source}} \Big),
\end{equation}

\noindent where $J$, $j$, and $\widetilde j$ are the sources, and
$M_I$ are masses of the Pauli--Villars fields.

Let us describe  all parts of this expression:

\begin{eqnarray}\label{Regularized_Action}
&& S_{\mbox{\scriptsize reg}} = \frac{1}{4e_0^2}\mbox{Re}\int
d^4x\,d^2\theta\,W^a R(\partial^2/\Lambda^2) W_a + \frac{1}{4}
\int d^4x\,d^4\theta\,\Big(\phi^* e^{2V}\phi +
\widetilde\phi^* e^{-2V} \widetilde\phi\Big);\qquad\\
\label{PV_Action} && S_{\mbox{\scriptsize gf}} = - \frac{1}{64
e_0^2}\int d^4x\,d^4\theta\, \Big(V R(\partial^{2}/\Lambda^{2})
D^2 \bar D^2
V + V R(\partial^{2}/\Lambda^{2}) \bar D^2 D^2 V\Big);\\
\label{Sources_Action} && S_{\mbox{\scriptsize source}} = \int
d^4x\,d^4\theta\,V J + \Big(\int d^4x\,d^2\theta\, (\phi j +
\widetilde\phi\, \widetilde j ) +\mbox{c.c.}\Big).
\end{eqnarray}

\noindent In the Abelian case it is not necessary to introduce
ghost (super)fields. The Pauli--Villars determinants can be
presented as functional integrals over the corresponding (chiral)
Pauli--Villars superfields $\Phi$ and $\widetilde\Phi$:

\begin{equation}
\Big(\det(V,M)\Big)^{-1} = \int
D\Phi\,D\widetilde\Phi\,\exp(iS_{\mbox{\scriptsize PV}}),
\end{equation}

\noindent where the action for the Pauli--Villars superfields is

\begin{equation}\label{PV_Action1}
S_{\mbox{\scriptsize PV}} = \frac{1}{4}\int d^8x\,\Big(\Phi^*
e^{2V}\Phi + \widetilde\Phi^* e^{-2V}\widetilde\Phi \Big) +
\Big(\frac{1}{2} \int d^4x\,d^2\theta\,M \Phi \widetilde\Phi
+\mbox{c.c.}\Big).
\end{equation}

\noindent For cancellation of remaining one-loop divergences the
coefficients $c_I$ should satisfy the conditions
\cite{Slavnov:1977zf}

\begin{equation}\label{C_Conditions}
\sum\limits_I c_I =1;\qquad \sum\limits_I c_I M_I^2 =0.
\end{equation}

\noindent We choose masses of the Pauli--Villars fields so that
they will be proportional to the parameter $\Lambda$ in the higher
derivative term \cite{Stepanyantz:2011wq,Stepanyantz:2011jy}:

\begin{equation}\label{Scheme}
M_I = a_I\Lambda,
\end{equation}

\noindent where $a_I$ are arbitrary real constants which do not
depend of the bare charge. Then the NSVZ relation is obtained
exactly in all orders for the RG functions defined in terms of the
bare coupling constant independently of a concrete renormalization
prescription. Let us briefly describe, how this can be  shown.

Due to the Ward identity 
the two-point Green  function of the gauge superfield is
transverse:

\begin{equation}\label{D_Definition}
\Gamma^{(2)}_{V} - S_{\mbox{\scriptsize gf}} = - \frac{1}{16\pi}
\int
\frac{d^4p}{(2\pi)^4}\,d^4\theta\,V(\theta,-p)\,\partial^2\Pi_{1/2}
V(\theta,p)\, d^{-1}(\alpha_0,\Lambda/p),
\end{equation}

\noindent where the supersymmetric transversal projector is given
by

\begin{equation}
\partial^2\Pi_{1/2}  = -\frac{1}{8} D^a \bar D^2 D_a.
\end{equation}

\noindent The function $d$ defined by Eq. (\ref{D_Definition})
coincides with the invariant charge  $\alpha_{\mbox{\scriptsize
inv}}$ \cite{Bogolyubov:1980nc}, related with the polarization
operator $\Pi(\alpha_0,\Lambda/p)$ by the equation

\begin{equation}
d_0^{-1}(\alpha_0, \Lambda/p) = \frac{1}{\alpha_0} \Big(1
+\frac{\alpha_0}{\pi} \Pi_0(\alpha_0, \Lambda/p)\Big)=\frac{1}
{\alpha_{\mbox{\scriptsize inv}}(\alpha_0,\Lambda/p)}.
\end{equation}

\noindent Similarly, a part of the effective action corresponding
to the two-point Green function of the matter superfields can be
written in the form

\begin{equation}\label{Gamma_Matter}
\Gamma^{(2)}_\phi = \frac{1}{4}\int \frac{d^4p}{(2\pi)^4}
d^4\theta\, \Big(\phi^*(\theta,-p)\, \phi(\theta,p) +
\widetilde\phi^*(\theta,-p)\, \widetilde\phi(\theta,p) \Big)
G(\alpha_0,\Lambda/p).
\end{equation}

Here we use the following renormalization procedure
\cite{Bogolyubov:1980nc,Collins:1984xc}: we define a renormalized
coupling constant $\alpha(\alpha_0,\Lambda/\mu)$, where $\mu$ is a
renormalization parameter, requiring that the inverse  invariant
charge $d^{-1}(\alpha_0(\alpha,\Lambda/\mu),\Lambda/p)$ is finite
in the limit $\Lambda\to \infty$ (and, therefore, in this limit
depends only on $\alpha$ and $\mu/p$). Then the renormalization
constant $Z_3$ is defined according to the equation

\begin{equation}
\frac{1}{\alpha_0} \equiv \frac{Z_3(\alpha,\Lambda/\mu)}{\alpha}.
\end{equation}

\noindent In order to renormalize the two-point Green function of
the matter superfields, we construct the renormalization constant
$Z$, requiring that the renormalized two-point Green function $Z
G$ is finite in the limit $\Lambda\to \infty$:

\begin{equation}
G_{\mbox{\scriptsize ren}}(\alpha,\mu/p) = \lim\limits_{\Lambda\to
\infty} Z(\alpha, \Lambda/\mu)G(\alpha_0,\Lambda/p).
\end{equation}

\noindent Certainly, the renormalized coupling constant $\alpha$
and the renormalization constant $Z$ are not uniquely defined and
depend on a choice of a renormalization scheme
\cite{Vladimirov:1975mx}.

The RG functions are usually defined in terms of the renormalized
coupling constant. This definition is presented in Sect.
\ref{Section_Boundary_Conditions}. However, it is also possible to
define the RG functions in terms of the bare coupling constant
$\alpha_0$ according to the following prescription:

\begin{eqnarray}\label{Beta_Definition1}
&& \beta\Big(\alpha_0(\alpha,\Lambda/\mu)\Big) \equiv \frac{d
\alpha_0(\alpha,\Lambda/\mu)}{d\ln\Lambda}
\Big|_{\alpha=\mbox{\scriptsize const}};\vphantom{\Bigg|}\\
\label{Gamma_Definition1} &&
\gamma\Big(\alpha_0(\alpha,\Lambda/\mu)\Big) \equiv - \frac{d \ln
Z(\alpha,\Lambda/\mu)}{d
\ln\Lambda}\Big|_{\alpha=\mbox{\scriptsize const}},
\end{eqnarray}

\noindent where $\alpha$ and $\Lambda$ are considered as
independent variables. Nevertheless, for finding these functions
it is necessary to use the relation between the bare coupling
constant $\alpha_0$ and the renormalized coupling constant
$\alpha$. Really, the differentiation with respect to $\Lambda$ is
made at a fixed value of $\alpha$. Therefore, before this
differentiating we should express $\alpha_0$ in terms of $\alpha$
and $\Lambda/\mu$. This implies that expressions dependent on the
renormalization scheme appear at intermediate steps of the
calculation. However, final expressions for the RG functions
(\ref{Beta_Definition1}) and (\ref{Gamma_Definition1}) are
independent of a renormalization prescription. In order to see
this, let us consider the function

\begin{equation}
d^{-1}(\alpha_0,\Lambda/p) = d^{-1}(\alpha,\mu/p) + \mbox{(terms
vanishing in the limit $p\to 0$)}
\end{equation}

\noindent and differentiate it with respect to $\ln\Lambda$ (at a
fixed value of $\alpha$). The derivative of the right hand side
evidently vanishes in the limit $p\to 0$. Therefore,

\begin{equation}\label{Vanishing_Derivative}
0 = \lim\limits_{p\to 0} \frac{d
d^{-1}(\alpha_0,\Lambda/p)}{d\ln\Lambda}
\Big|_{\alpha=\mbox{\scriptsize const}} = \lim\limits_{p\to
0}\Big(\frac{\partial d^{-1}(\alpha_0,\Lambda/p)}{\partial
\alpha_0} \beta(\alpha_0) - \frac{\partial
d^{-1}(\alpha_0,\Lambda/p)}{\partial\ln p} \Big),
\end{equation}

\noindent where in the last equality $\alpha_0$ and $p$ are
considered as independent variables. From this equation the
$\beta$-function (\ref{Beta_Definition1}) is expressed through the
expressions which evidently do not depend on a renormalization
prescription. Therefore, this function does not depend on a choice
of a renormalization scheme. Similarly, differentiating the
equality

\begin{equation}
\ln G(\alpha_0,\Lambda/q) = \ln G_{\mbox{\scriptsize
ren}}(\alpha,\mu/q) - Z(\alpha,\Lambda/\mu) + \mbox{(terms
vanishing in the limit $q\to 0$)}
\end{equation}

\noindent with respect to $\ln\Lambda$ at a fixed value of
$\alpha$, in the limit $q\to 0$ we obtain

\begin{equation}
\gamma(\alpha_0) = \lim\limits_{q\to 0} \Big(\frac{\partial\ln
G(\alpha_0,\Lambda/q)}{\partial \alpha_0} \beta(\alpha_0) -
\frac{\partial\ln G(\alpha_0,\Lambda/q)}{\partial \ln q}\Big).
\end{equation}

\noindent Therefore, the anomalous dimension
(\ref{Gamma_Definition1}) does not depend on a choice of the
function $\alpha_0(\alpha,\Lambda/\mu)$ as well.

In order to find an expression for the $\beta$-function
(\ref{Beta_Definition1}) in the case of ${\cal N}=1$ SQED
regularized by higher derivatives, it is possible to make the
substitution

\begin{equation}\label{V_Substitution}
V \to \bar \theta^a \bar\theta_a \theta^b \theta_b\equiv \theta^4
\end{equation}

\noindent in the expression

\begin{equation}
\Delta\Gamma^{(2)}_V = \Gamma^{(2)}_V - S_{\mbox{\scriptsize gf}}
- S.
\end{equation}

\noindent Then using Eq. (\ref{D_Definition}) we obtain

\begin{eqnarray}\label{Calculate2}
&& \frac{d(\Delta\Gamma^{(2)}_{\mbox{\scriptsize
V})}}{d\ln\Lambda} \Big|_{V(x,\theta)=\theta^4} =
(2\pi)^3\delta^{4}(p) \frac{d}{d\ln \Lambda}\,
\Big(d^{-1}(\alpha_0,\Lambda/p)-\alpha_0^{-1}\Big)\nonumber\\
&&\qquad\qquad\qquad\qquad\qquad\qquad\qquad\qquad = -
(2\pi)^3\delta^{4}(p) \frac{d\alpha_0^{-1}}{d \ln \Lambda} =
(2\pi)^3\delta^{4}(p) \frac{\beta(\alpha_0)}{\alpha_0^2},\qquad
\end{eqnarray}

\noindent where (exactly as in Eqs. (\ref{Beta_Definition1}) and
(\ref{Gamma_Definition1})) $\alpha$ and $\Lambda$ are considered
as independent variables, and we take into account the first
equality in Eq. (\ref{Vanishing_Derivative}).\footnote{The limit
$p\to 0$, in which the derivative of the function $d^{-1}$
vanishes, is obtained due to the factor $\delta^4(p)$.}

The crucial observation is that if the higher derivative method is
used for a regularization, the integrals obtained by calculating
the left hand side of Eq. (\ref{Calculate2}) are integrals of
(double) total derivatives
\cite{Soloshenko:2003nc,Smilga:2004zr,Stepanyantz:2011jy}.
Therefore, one of the loop integrals can be calculated
analytically. Using a rather complicated technique
\cite{Stepanyantz:2011jy}, it is possible to obtain

\begin{equation}\label{Green_Function_Relation}
\frac{\beta(\alpha_0)}{\alpha_0^2} = \frac{d}{d\ln \Lambda}\,
\Big(d^{-1}(\alpha_0,\Lambda/p)-\alpha_0^{-1}\Big)\Big|_{p=0} =
\frac{1}{\pi}\Big(1-\frac{d}{d\ln\Lambda} \ln
G(\alpha_0,\Lambda/q)\Big|_{q=0}\Big)
\end{equation}

\noindent exactly in all orders. The right hand side of this
equation can be expressed through the anomalous dimension
(\ref{Gamma_Definition1}):

\begin{eqnarray}\label{NSVZ_Bare}
&& \frac{1}{\pi}\Big(1-\frac{d}{d\ln\Lambda} \ln
G(\alpha_0,\Lambda/q)\Big|_{q=0}\Big) = \frac{1}{\pi}-
\frac{1}{\pi}\frac{d}{d\ln\Lambda} \Big(\ln G_{\mbox{\scriptsize
ren}}(\alpha,\mu/q) - \ln
Z(\alpha,\Lambda/\mu) \Big)\Big|_{q=0}\nonumber\\
&& = \frac{1}{\pi} \Big(1 -
\gamma\Big(\alpha_0(\alpha,\Lambda/\mu)\Big)\Big),
\end{eqnarray}

\noindent where we take into account that the function $Z G$ does
not depend on $\Lambda$ in the limit $q\to 0$. Thus, we obtain

\begin{equation}
\frac{\beta(\alpha_0)}{\alpha_0^2}= \frac{1}{\pi} \Big(1 -
\gamma\Big(\alpha_0(\alpha,\Lambda/\mu)\Big)\Big).
\end{equation}

\noindent This relation gives the NSVZ $\beta$-function
(\ref{NSVZ}), which was first obtained in
\cite{Vainshtein:1986ja,Shifman:1985fi} using a different method.

It is important that this relation between the RG functions
defined in terms of the {\it bare} coupling constant is valid for
an arbitrary choice of the functions
$\alpha_0(\alpha,\Lambda/\mu)$ and $Z(\alpha,\Lambda/\mu)$, which
specify a  subtraction scheme. (Note  that at intermediate steps
these functions are needed for calculating the RG functions
(\ref{Beta_Definition1}) and (\ref{Gamma_Definition1}).) Thus,
with the higher derivative regularization the NSVZ
$\beta$-function is naturally obtained in terms of the bare
coupling constant.

\section{Regularization dependence
of the RG functions defined in terms of the  bare charge}
\hspace{\parindent} \label{Section_Rescaling}

The RG functions (\ref{Beta_Definition1}) and
(\ref{Gamma_Definition1}) are related by Eq. (\ref{NSVZ})
independently of a particular choice of counterterms, if the
theory is regularized according to the prescription described in
Sect. \ref{Section_Regularization}. However, it is possible to
change the higher derivative regularization so that to modify the
NSVZ relation. This can be done, for example, by a bare coupling
constant dependent rescaling of the Pauli--Villars masses. Let us
investigate, how the NSVZ relation for bare quantities is modified
after this rescaling.

Actually, the regularization for which the NSVZ relation is
obtained for the RG functions (\ref{Beta_Definition1}) and
(\ref{Gamma_Definition1}) is singled out by the condition
(\ref{Scheme}). More exactly, the coefficients $a_I$ should not
depend on $\alpha_0$. However, one-loop divergences can be also
cancelled, if we choose

\begin{equation}\label{MZ}
M_I = a_I \Lambda/z_0(\alpha_0),
\end{equation}

\noindent where $z(\alpha_0)$ is an arbitrary finite function of
the bare coupling constant $\alpha_0 = e_0^2/4\pi$. This is
equivalent to the redefinition of the $S_{\mbox{\scriptsize reg}}$
and $S_{\mbox{\scriptsize PV}}$:

\begin{eqnarray}\label{Z_Regularized_Action}
&& S_{\mbox{\scriptsize reg}} = \frac{1}{4e_0^2}\mbox{Re}\int
d^4x\,d^2\theta\,W^a W_a + \frac{1}{4}z_0(\alpha_0) \int
d^4x\,d^4\theta\,\Big(\phi^* e^{2V}\phi + \widetilde\phi^* e^{-2V}
\widetilde\phi\Big);\\
\label{Z_PV_Action} && S_{\mbox{\scriptsize PV}} =
\frac{1}{4}z_0(\alpha_0) \int d^8x\,\Big(\Phi^* e^{2V}\Phi +
\widetilde\Phi^* e^{-2V}\widetilde\Phi \Big) + \Big(\frac{1}{2}
\int d^4x\,d^2\theta\,M \Phi \widetilde\Phi
+\mbox{c.c.}\Big).\qquad
\end{eqnarray}

\noindent These actions are obtained after rescaling of the
intergration variables

\begin{equation}
\phi\to \sqrt{z_0(\alpha_0)}\,\phi;\qquad \widetilde\phi\to
\sqrt{z_0(\alpha_0)}\, \widetilde\phi;\qquad \Phi_I \to
\sqrt{z_0(\alpha_0)}\,\Phi_I;\qquad \widetilde\Phi_I \to
\sqrt{z_0(\alpha_0)}\, \widetilde\Phi_I.
\end{equation}

\noindent in the generating functional
(\ref{Generating_Functional}). In order to avoid anomalous
contributions of the integration measure, discussed in
\cite{ArkaniHamed:1997mj}, the Pauli--Villars fields should be
rescaled in the same way as the usual fields. Then
$S_{\mbox{\scriptsize reg}}$ given by Eq.
(\ref{Z_Regularized_Action}) and $S_{\mbox{\scriptsize PV}}$ given
by Eq. (\ref{Z_PV_Action}) produce diagrams with  insertions of

\begin{equation}
\frac{1}{4}(z_0(\alpha_0)-1) \int d^4x\,d^4\theta\,\Big(\phi^*
e^{2V}\phi + \widetilde\phi^* e^{-2V} \widetilde\phi\Big)
\end{equation}

\noindent on the matter lines and similar diagrams with closed
loops of the Pauli--Villars fields.  In order to calculate a
contribution of these diagrams, we note that Eq.
(\ref{Green_Function_Relation}) is also valid for all diagrams
beyond the one-loop approximation and for all values of $z_0$.
This can be proved repeating the calculations made in
\cite{Stepanyantz:2011jy} taking into account that dependence on
the Pauli--Villars masses is the same in both sides of this
equation. However, a one-loop contribution to the right hand side
of Eq. (\ref{Green_Function_Relation}) is

\begin{equation}
\label{35} \frac{1}{\pi}\cdot \frac{d}{d\ln\Lambda}
\sum\limits_{I=1}^n c_I \ln \frac{M_I}{p}\Big|_{p=0}.
\end{equation}

\noindent If the Pauli--Villars masses are given by Eq.
(\ref{MZ}), instead of Eq. (\ref{35}) we obtain

\begin{equation}
\frac{1}{\pi} \sum\limits_{I=1}^n c_I \Big(1-\frac{d\ln
z_0}{d\ln\Lambda}\Big) = \frac{1}{\pi} \Big(1-\frac{d\ln
z_0}{d\ln\Lambda}\Big) = \frac{1}{\pi} \Big(1-\beta(\alpha_0)
\frac{d\ln z_0}{d\alpha_0}\Big).
\end{equation}

\noindent For $z_0=1$ this gives  $1/\pi$ in the right hand side
of Eq. (\ref{Green_Function_Relation}). Now it is evident that for
an arbitrary function $z_0$ the generalization of Eq.
(\ref{Green_Function_Relation}) can be  written as

\begin{equation}
\frac{d}{d\ln\Lambda} \Big(d^{-1}(\alpha_0,\Lambda/p) -
\alpha_0^{-1}\Big)\Big|_{p=0}  =
\frac{1}{\pi}\Big(1-\frac{d}{d\ln\Lambda} \ln
G(\alpha_0,\Lambda/q)\Big|_{q=0} - \frac{d\ln
z_0(\alpha_0)}{d\ln\Lambda} \Big).
\end{equation}

\noindent From this equation we easily obtain that the
$\beta$-function (\ref{Beta_Definition1}), which depends on the
bare coupling constant $\alpha_0$,  has the form

\begin{equation}
\beta(\alpha_0) = \frac{\alpha_0^2}{\pi} \cdot \frac{1 -
\gamma(\alpha_0)}{1 + \alpha_0^2\, (d\ln z_0/d\alpha_0)/\pi}.
\end{equation}

\noindent Therefore, changing the Pauli--Villars masses according to
Eq. (\ref{MZ}) we modify the expression for the exact
$\beta$-function expressed in terms of the bare coupling constant.

Also the NSVZ relation can be modified  by a finite tuning of the
bare charge $\alpha_0 \to \alpha_0'(\alpha_0)$. Combining this
finite tuning with the rescaling of the Pauli--Villars masses, it
is possible to obtain the RG functions dependent on the bare
charge with coefficients equal to those obtained within the
$\overline{\mbox{DR}}$ approach.

\section{NSVZ scheme}
\hspace{\parindent}\label{Section_Boundary_Conditions}

Although the above prescriptions allow to obtain the NSVZ
$\beta$-function in terms of the bare coupling constant, it is
possible to study a problem  whether  the NSVZ relation is valid
for the RG functions defined in terms of the renormalized coupling
constant, if the theory is regularized by higher derivatives.

Let us consider the RG functions, defined in terms of the
renormalized coupling constant:

\begin{eqnarray}
\label{Beta_Definition2} &&
\widetilde\beta\Big(\alpha(\alpha_0,\Lambda/\mu)\Big) \equiv
\frac{d\alpha(\alpha_0,\Lambda/\mu)}{d\ln\mu}\Big|_{\alpha_0=\mbox{\scriptsize const}};\\
\label{Gamma_Definition2} &&
\widetilde\gamma\Big(\alpha(\alpha_0,\Lambda/\mu)\Big) \equiv
\frac{d}{d\ln\mu}\ln
ZG(\alpha_0,\Lambda/\mu)\Big|_{\alpha_0=\mbox{\scriptsize const}}
= \frac{d\ln Z(\alpha(\alpha_0,\Lambda/\mu),
\Lambda/\mu)}{d\ln\mu}\Big|_{\alpha_0=\mbox{\scriptsize
const}},\qquad
\end{eqnarray}

\noindent where $\alpha_0$ and $\mu$ are considered as independent
variables. These functions depend on the arbitrariness of choosing
$\alpha$ and $Z$. As a consequence, in general, these functions
do not satisfy the NSVZ relation, which was originally derived for
the bare quantities.

The arbitrariness of choosing $\alpha$ and $Z$ can be fixed by
imposing certain boundary conditions. For example, the momentum
subtraction (MOM) scheme is defined by the following
requirements:

\begin{equation}\label{MOM_Boundary_Conditions}
Z_{\mbox{\scriptsize MOM}}G(\alpha_{\mbox{\scriptsize MOM}},p=\mu)
= 1; \qquad d^{-1}(\alpha_{\mbox{\scriptsize MOM}},p=\mu)
=\alpha_{\mbox{\scriptsize MOM}}^{-1}.
\end{equation}

\noindent In terms of the polarization operator the second
condition can be rewritten as
$\Pi(\alpha_{\mbox{\scriptsize MOM}},p=\mu) = 0$. In the MOM
scheme the $\beta$-function (\ref{Beta_Definition2}) coincides
with the Gell-Mann--Low function $\Psi(\alpha)$ (see
\cite{Gorishnii:1990kd} for the detailed explanation).

The other schemes can be obtained by making finite renormalizations:

\begin{equation}\label{Finite_Renormalization}
\frac{1}{\alpha} = \frac{z_3(\alpha_{\mbox{\scriptsize
MOM}})}{\alpha_{\mbox{\scriptsize MOM}}};\qquad
Z(\alpha,\Lambda/\mu) = z(\alpha_{\mbox{\scriptsize MOM}})\,
Z_{\mbox{\scriptsize MOM}}(\alpha_{\mbox{\scriptsize
MOM}},\Lambda/\mu),
\end{equation}

\noindent where $z_3$ and $z$ are finite functions. In the
three-loop approximation such a  renormalization  relating the
NSVZ scheme and the MOM scheme is constructed in Appendix
\ref{Appendix_Renormalization}.

In this section we construct a scheme of the ultraviolet
renormalization for which the NSVZ relation is also satisfied for
the renormalized functions in case of using the higher derivative
regularization. This scheme is related with the MOM scheme by the
finite renormalization (\ref{Finite_Renormalization}). It is
constructed by imposing an additional condition, which can be
formulated as follows: There should be a point $x_0
=\ln\Lambda/\mu_0$ such that

\begin{equation}\label{NSVZ_Scheme}
\alpha_0(\alpha_{\mbox{\scriptsize NSVZ}},x_0) =
\alpha_{\mbox{\scriptsize NSVZ}}; \qquad Z_{\mbox{\scriptsize
NSVZ}}(\alpha_{\mbox{\scriptsize NSVZ}},x_0)=1
\end{equation}

\noindent for all values of $\alpha_{\mbox{\scriptsize NSVZ}}$.
Equivalently, there is a point $x_0$ such that

\begin{equation}\label{NSVZ_Scheme_Z3}
(Z_3)_{\mbox{\scriptsize NSVZ}}(\alpha_{\mbox{\scriptsize
NSVZ}},x_0) = 1;\qquad Z_{\mbox{\scriptsize
NSVZ}}(\alpha_{\mbox{\scriptsize NSVZ}},x_0)=1.
\end{equation}

Although these conditions  are similar to the requirement $Z_3=1$,
which was used in Ref. \cite{Kataev:2013vua} to define the
conformally invariant limit of perturbative QED \cite{Kataev:2010tm},
Eqs. (\ref{NSVZ_Scheme}) and
(\ref{NSVZ_Scheme_Z3}) are imposed in a single point $x_0$ only.
In Appendix \ref{Appendix_Existence} we prove that the scheme
defined by Eq. (\ref{NSVZ_Scheme}) exists and without loss of
generality it is possible to choose $x_0=0$. This result is also
verified by an explicit calculation in the three-loop
approximation.

Let us prove this statement. We assume that Eq.
(\ref{NSVZ_Scheme}) is satisfied. It is convenient to define a
variable $x = \ln\Lambda/\mu$. Then the RG functions
(\ref{Beta_Definition2}) and (\ref{Gamma_Definition2}) can be written as

\begin{eqnarray}
&& \widetilde\beta\left(\alpha(\alpha_0,x)\right) =
-\frac{\partial
\alpha(\alpha_0,x)}{\partial x}; \nonumber\\
&& \widetilde\gamma\left(\alpha(\alpha_0,x)\right) = -\frac{d \ln
Z\left(\alpha(\alpha_0,x),x\right)}{d x} = - \frac{\partial \ln
Z(\alpha,x)}{\partial \alpha}\cdot
\frac{\partial\alpha(\alpha_0,x)}{\partial x} - \frac{\partial \ln
Z\left(\alpha(\alpha_0,x),x\right)}{\partial x},\qquad
\end{eqnarray}

\noindent where the total derivative with respect to $x$ also acts
on $x$ inside $\alpha$. Calculating these expressions at the point
$x=x_0$ and taking into account Eq. (\ref{NSVZ_Scheme}) we obtain

\begin{equation}\label{Equality}
\widetilde\beta(\alpha_0) = \beta(\alpha_0);\qquad
\widetilde\gamma(\alpha_0) = \gamma(\alpha_0).
\end{equation}

\noindent Really, in order to prove the first equality we note
that

\begin{eqnarray}
&& \alpha_0(\alpha,x) = \alpha + \beta(\alpha_0) (x-x_0) +
O\left((x-x_0)^2\right);\nonumber\\
&& \alpha(\alpha_0,x) = \alpha_0 - \widetilde\beta(\alpha) (x-x_0)
+ O\left((x-x_0)^2\right)\nonumber\\
&&\qquad\qquad\qquad\qquad\qquad = \alpha_0 -
\widetilde\beta(\alpha_0) (x-x_0) + O\left((x-x_0)^2\right).\qquad
\end{eqnarray}

\noindent Similarly, the second equation in (\ref{Equality}) can
be obtained taking into account that

\begin{equation}
\frac{\partial \ln Z(\alpha,x_0)}{\partial \alpha} = 0
\end{equation}

\noindent due to Eq. (\ref{NSVZ_Scheme}). According to
\cite{Stepanyantz:2011jy} the functions (\ref{Beta_Definition1})
and (\ref{Gamma_Definition1}) satisfy the equation

\begin{equation}
\beta(\alpha_0) =
\frac{\alpha_0^2}{\pi}\Big(1-\gamma(\alpha_0)\Big),
\end{equation}

\noindent if one uses the regularization described in Sect.
\ref{Section_Regularization}. Therefore, from Eq. (\ref{Equality})
we conclude that

\begin{equation}
\widetilde\beta(\alpha) =
\frac{\alpha^2}{\pi}\Big(1-\widetilde\gamma(\alpha)\Big),
\end{equation}

\noindent if the boundary conditions (\ref{NSVZ_Scheme}) are
imposed. In the three-loop approximation this is verified by an
explicit calculation in Sect. \ref{Section_Explicit_Three_Loop}.

\section{The scheme dependence in the three-loop approximation}
\hspace{\parindent}\label{Section_Explicit_Three_Loop}

Let us reconsider the results of the three-loop calculation
\cite{Soloshenko:2003nc} for ${\cal N}=1$ SQED and investigate the
scheme dependence of the RG functions in this approximation. (The
details of the three-loop calculation are described in Appendix
\ref{Appendix_Renormalization}.) We will start with  the two-point
Green function of the matter superfields. If the Pauli--Villars
masses are chosen according to the prescription (\ref{Scheme}), in
the two-loop approximation this function is given by the following
expression \cite{Soloshenko:2003sx}:\footnote{The expression
presented in \cite{Soloshenko:2003sx} is slightly different,
because it includes diagrams with one-loop counterterm insertions
on the matter lines and, therefore, corresponds to
$Z_{\mbox{\scriptsize 1-loop}} G_{\mbox{\scriptsize 2-loop}}$.}

\begin{eqnarray}\label{G(alpha0)}
&& G(\alpha_0,\Lambda/p) = 1 - \int \frac{d^4k}{(2\pi)^4}
\frac{2e_0^2}{k^2 R_k (k+p)^2} + \int \frac{d^4k}{(2\pi)^4}
\frac{d^4l}{(2\pi)^4} \frac{4e_0^4}{k^2 R_k l^2 R_l}
\Bigg(\frac{1}{(k+p)^2 (l+p)^2}\nonumber\\
&& + \frac{1}{(l+p)^2 (k+l+p)^2} - \frac{(k+l+2p)^2}{(k+p)^2
(l+p)^2 (k+l+p)^2} \Bigg) + \int \frac{d^4k}{(2\pi)^4}
\frac{d^4l}{(2\pi)^4} \frac{4 e_0^4}{k^2 R_k^2 (k+p)^2}
\nonumber\\
&& \times \Bigg(\frac{1}{l^2(k+l)^2} - \sum\limits_{I=1}^n c_I
\frac{1}{\left(l^2 + M_I^2\right)\left((k+l)^2+M_I^2\right)}\Bigg)
+ O(e_0^6) ,
\end{eqnarray}

\noindent where $R_k\equiv R(k^2/\Lambda^2)$. (This expression is
written in the Euclidean space after the Wick rotation.)

We split the bare coupling constant $\alpha_0 = e_0^2/4\pi$ into
the renormalized coupling constant $\alpha$ and a counterterm. In
the lowest approximation

\begin{equation}\label{One-Loop_Coupling_Renormalization}
\frac{1}{\alpha_0} = \frac{1}{\alpha} - \frac{1}{\pi} \Big(
\ln\frac{\Lambda}{\mu} + b_1\Big) + O(\alpha),
\end{equation}

\noindent where the logarithm compensates a one-loop divergence
and a finite parameter $b_1$ is not fixed. Different choices of
$b_1$ correspond to different subtraction schemes. Equivalently,
Eq. (\ref{One-Loop_Coupling_Renormalization}) can be rewritten in
the   form

\begin{equation}
\alpha_0 = \alpha\Big(1 + \frac{\alpha}{\pi}
\Big(\ln\frac{\Lambda}{\mu} + b_1\Big) + O(\alpha^2) \Big).
\end{equation}

In order to cancel divergences in the two-point Green function
of the matter superfields, the function $G(\alpha_0,\Lambda/p)$
should be multiplied by the renormalization constant $Z$. This
corresponds to the following redefinition of the matter
superfields:

\begin{equation}\label{Rescaling_Of_Arguments}
\phi = Z^{1/2}(\alpha,\Lambda/\mu)\,\phi_{\mbox{\scriptsize
ren}};\qquad \widetilde \phi =
Z^{1/2}(\alpha,\Lambda/\mu)\,\widetilde\phi_{\mbox{\scriptsize
ren}},
\end{equation}

\noindent where $\phi$ and $\widetilde\phi$ are the bare fields,
and $\phi_{\mbox{\scriptsize ren}}$ and
$\widetilde\phi_{\mbox{\scriptsize ren}}$ are the renormalized
fields.  It is easy to see that in the one-loop approximation the
renormalization constant $Z$, which is chosen so that the
renormalized Green function $Z G$ should be finite,  is

\begin{equation}
Z = 1 + \frac{\alpha}{\pi}\Big(\ln\frac{\Lambda}{\mu}+g_1\Big) +
O(\alpha^2),
\end{equation}

\noindent where $g_1$ is an arbitrary finite constant. In
principle, without loss of generality the constant $g_1$ can be
excluded by a redefinition of the parameter $\mu$. However, we
will not do this, because in the next orders similar coefficients
cannot be excluded by this way. Choosing a higher derivative term
in the momentum space as

\begin{equation}\label{Explicit_Regulator}
R_k = 1+\frac{k^{2n}}{\Lambda^{2n}},
\end{equation}

\noindent it is possible to find the following explicit expression
for the constant $Z$ in the next order \cite{Soloshenko:2003sx}:

\begin{equation}\label{Two_Loop_Z}
Z = 1 + \frac{\alpha}{\pi}\Big(\ln\frac{\Lambda}{\mu}+g_1\Big)
+\frac{\alpha^2}{\pi^2}\ln^2\frac{\Lambda}{\mu}
-\frac{\alpha^2}{\pi^2}\ln\frac{\Lambda}{\mu}\Big(\sum\limits_{I=1}^n
c_I\ln a_I - b_1 + \frac{3}{2} - g_1\Big) + \frac{\alpha^2
g_2}{\pi^2} + O(\alpha^3),
\end{equation}

\noindent where

\begin{equation}
a_I = M_I/\Lambda
\end{equation}

\noindent and $g_2$ is another finite constant. Choosing $g_2$ we
partially fix the subtraction scheme in the considered
approximation. The subtraction scheme in the considered
approximation is completely fixed by choosing $g_1$, $g_2$, and the
finite constants in the renormalization of the coupling constant,
$b_1$ and $b_2$ (defined below).

Now, let us calculate the anomalous dimension according to Eqs.
(\ref{Gamma_Definition1}) and (\ref{Gamma_Definition2}).

In order to find the anomalous dimension according to the
prescription (\ref{Gamma_Definition1}), we should differentiate

\begin{eqnarray}
&& \ln Z(\alpha,\Lambda/\mu) =
\frac{\alpha}{\pi}\Big(\ln\frac{\Lambda}{\mu}+g_1\Big)
+\frac{\alpha^2}{2\pi^2}\ln^2\frac{\Lambda}{\mu} \nonumber\\
&&\qquad\qquad\qquad\qquad\quad\ \
-\frac{\alpha^2}{\pi^2}\ln\frac{\Lambda}{\mu}\Big(\sum\limits_{I=1}^n
c_I\ln a_I - b_1 + \frac{3}{2}\Big) + \frac{\alpha^2}{\pi^2}
\Big(g_2-\frac{1}{2}g_1^2\Big) + O(\alpha^3)\qquad
\end{eqnarray}

\noindent with respect to $\ln\Lambda$ and write the result in
terms of $\alpha_0$. Then we obtain

\begin{eqnarray}\label{Gamma_Answer1}
&& \gamma(\alpha_0) = - \frac{d\ln Z}{d\ln\Lambda} =
-\frac{\alpha}{\pi} -
\frac{\alpha^2}{\pi^2}\Big(\ln\frac{\Lambda}{\mu} -
\sum\limits_{I=1}^n c_I \ln a_I + b_1 -
\frac{3}{2}\Big) + O(\alpha^3)\nonumber\\
&&\qquad\qquad\qquad\qquad\qquad\qquad\qquad\qquad\qquad =
-\frac{\alpha_0}{\pi} +
\frac{\alpha_0^2}{\pi^2}\Big(\frac{3}{2}+\sum\limits_{I=1}^n c_I
\ln a_I \Big) + O(\alpha_0^3).\qquad
\end{eqnarray}

\noindent This expression is independent of the finite constants
$g_i$ and $b_i$, which fix  the subtraction scheme.

The anomalous dimension $\widetilde\gamma(\alpha)$ defined by Eq.
(\ref{Gamma_Definition2}) can be constructed similarly. For this
purpose we rewrite $\ln Z$ in terms of $\alpha_0$ using Eq.
(\ref{One-Loop_Coupling_Renormalization}):

\begin{eqnarray}
&& \ln Z(\alpha(\alpha_0,\Lambda/\mu),\Lambda/\mu) =
\frac{\alpha_0}{\pi}\Big(\ln\frac{\Lambda}{\mu}+g_1\Big)
-\frac{\alpha_0^2}{2\pi^2}\ln^2\frac{\Lambda}{\mu} \nonumber\\
&&\qquad\qquad\qquad
-\frac{\alpha_0^2}{\pi^2}\ln\frac{\Lambda}{\mu}\Big(\sum\limits_{I=1}^n
c_I\ln a_I +g_1 + \frac{3}{2}\Big) + \frac{\alpha_0^2}{\pi^2}
\Big(g_2-\frac{1}{2}g_1^2 - g_1 b_1\Big) + O(\alpha_0^3).\qquad
\end{eqnarray}

\noindent Differentiating this expression with respect to $\ln\mu$
and writing the result in terms of $\alpha$ we obtain

\begin{eqnarray}\label{Gamma_Answer2}
&& \widetilde\gamma(\alpha) = \frac{d\ln Z}{d\ln\mu} =
-\frac{\alpha_0}{\pi} +
\frac{\alpha_0^2}{\pi^2}\Big(\ln\frac{\Lambda}{\mu} +
\sum\limits_{I=1}^n c_I \ln a_I + g_1 +
\frac{3}{2}\Big) + O(\alpha_0^3)\nonumber\\
&&\qquad\qquad\qquad\qquad\qquad\qquad\quad  = -\frac{\alpha}{\pi}
+ \frac{\alpha^2}{\pi^2}\Big(\frac{3}{2}+\sum\limits_{I=1}^n c_I
\ln a_I  - b_1 + g_1\Big) + O(\alpha^3).\qquad
\end{eqnarray}

\noindent Unlike Eq. (\ref{Gamma_Answer1}), this expression depends
on the constants $g_1$ and $b_1$.

Let us study, under what  conditions the NSVZ relation is valid.
For this purpose we write the explicit expression for the
three-loop renormalized coupling constant. The integrals defining
the three-loop $\beta$-function are presented, for example, in
Ref. \cite{Stepanyantz:2011jy}. They were obtained in
\cite{Soloshenko:2003nc}, but a more compact form for writing them
was proposed later.  If the higher derivative term is given by Eq.
(\ref{Explicit_Regulator}), these integrals can be calculated
explicitly using the results of \cite{Soloshenko:2003sx}. After
this calculation we split the bare coupling constant into the
renormalized coupling constant and a counterterm in such a way
that the function $d^{-1}$ is finite. It is possible to find (see
Appendix \ref{Appendix_Renormalization}) that the result can be
written in the following form:

\begin{eqnarray}\label{Alpha0}
&& \frac{1}{\alpha_0} = \frac{1}{\alpha} - \frac{1}{\pi} \Big(
\ln\frac{\Lambda}{\mu} + b_1\Big) - \frac{\alpha}{\pi^2}
\Big(\ln\frac{\Lambda}{\mu} + b_2\Big)\nonumber\\
&& \qquad\qquad\qquad -
\frac{\alpha^2}{\pi^3}\Big(\frac{1}{2}\ln^2 \frac{\Lambda}{\mu}
-\ln \frac{\Lambda}{\mu} \sum\limits_{I=1}^n c_I \ln a_I
-\frac{3}{2} \ln \frac{\Lambda}{\mu} + b_1 \ln\frac{\Lambda}{\mu}
+ b_3\Big) + O(\alpha^3),\qquad
\end{eqnarray}

\noindent where $b_1$, $b_2$, and $b_3$ are finite constants,
which define the subtraction scheme. (This three-loop  expression
coincides with Eq. (\ref{One-Loop_Coupling_Renormalization}) at
the one-loop level.)

Differentiating Eq. (\ref{Alpha0}) with respect to $\ln\Lambda$
and writing the result in terms of $\alpha_0$ we obtain the
$\beta$-function defined by Eq. (\ref{Beta_Definition1}):

\begin{eqnarray}\label{Beta_Expression}
&& \frac{\beta(\alpha_0)}{\alpha_0^2} = \frac{1}{\pi} +
\frac{\alpha}{\pi^2}  + \frac{\alpha^2}{\pi^3}\Big(\ln
\frac{\Lambda}{\mu} - \sum\limits_{I=1}^n c_I \ln a_I -\frac{3}{2}
+ b_1\Big) +
O(\alpha^3)\nonumber\\
&& \qquad\qquad\qquad\qquad\qquad\qquad\qquad\qquad =
\frac{1}{\pi} + \frac{\alpha_0}{\pi^2} -
\frac{\alpha_0^2}{\pi^3}\Big( \sum\limits_{I=1}^n c_I \ln a_I
+\frac{3}{2} \Big) + O(\alpha_0^3).\qquad
\end{eqnarray}

\noindent Thus, this $\beta$-function does not depend  on the
finite constants $g_i$ and $b_i$, as was proved in Eq.
(\ref{Vanishing_Derivative}). Moreover, comparing this expression
with Eq. (\ref{Gamma_Answer1}) we see that for all values of $g_i$
and $b_i$ the NSVZ relation is valid.

In order to calculate the $\beta$-function defined by Eq.
(\ref{Beta_Definition2}), we re-express $\alpha$ in terms of
$\alpha_0$:

\begin{eqnarray}\label{Alpha}
&& \frac{1}{\alpha} = \frac{1}{\alpha_0} + \frac{1}{\pi} \Big(
\ln\frac{\Lambda}{\mu} + b_1\Big) + \frac{\alpha_0}{\pi^2}
\Big(\ln\frac{\Lambda}{\mu} + b_2\Big)\nonumber\\
&& \qquad + \frac{\alpha_0^2}{\pi^3}\Big(-\frac{1}{2}\ln^2
\frac{\Lambda}{\mu} -\ln \frac{\Lambda}{\mu} \sum\limits_{I=1}^n
c_I \ln a_I -\frac{3}{2} \ln \frac{\Lambda}{\mu}
-b_2\ln\frac{\Lambda}{\mu} - b_1 b_2 + b_3\Big) +
O(\alpha_0^3).\qquad
\end{eqnarray}

\noindent Differentiating this expression with respect to $\ln
\mu$ and writing the result in terms of $\alpha$, we obtain

\begin{eqnarray}\label{Beta_Answer2}
&& \frac{\widetilde\beta(\alpha)}{\alpha^2} = \frac{1}{\pi} +
\frac{\alpha_0}{\pi^2}  + \frac{\alpha_0^2}{\pi^3}\Big(-\ln
\frac{\Lambda}{\mu} - \sum\limits_{I=1}^n c_I \ln a_I -\frac{3}{2}
-b_2\Big) +
O(\alpha_0^3)\nonumber\\
&& \qquad\qquad\qquad\qquad\qquad\quad\ \ = \frac{1}{\pi} +
\frac{\alpha}{\pi^2} - \frac{\alpha^2}{\pi^3}\Big(
\sum\limits_{I=1}^n c_I \ln a_I +\frac{3}{2} - b_1 + b_2\Big) +
O(\alpha^3).\qquad
\end{eqnarray}

\noindent In this case the $\beta$-function is related with the
anomalous dimension of the matter superfields
(\ref{Gamma_Answer2}) by Eq. (\ref{NSVZ}) only if finite constants
defining a subtraction scheme satisfy the condition

\begin{equation}
b_2 = g_1.
\end{equation}

\noindent It is easy to see that this identity  follows from the
system of equations

\begin{equation}\label{Three_Loop_Scheme}
\alpha_0 = \alpha(\alpha_0,\Lambda/\mu_0) + O(\alpha_0^4);\qquad 1
= Z(\alpha_0,\Lambda/\mu_0) + O(\alpha_0^2),
\end{equation}

\noindent which in the considered approximation coincides with Eq.
(\ref{NSVZ_Scheme}). Indeed, from the second equation we get

\begin{equation}
\ln\frac{\Lambda}{\mu_0} = -g_1.
\end{equation}

\noindent Substituting this relation to the first equation of
(\ref{Three_Loop_Scheme}) one can see that $b_1 = g_1$ and $b_2 =
g_1$. Although the constant $b_1$ is also fixed by Eq.
(\ref{Three_Loop_Scheme}), it is evident that the relation between
the $\beta$-function and the anomalous dimension is independent of
this constant. For $b_1=g_1$, $b_2=g_1$ in the considered
approximation $\gamma(\alpha) = \widetilde\gamma(\alpha)$ and
$\beta(\alpha) = \widetilde\beta(\alpha)$. This confirms the
results obtained in Sect. \ref{Section_Boundary_Conditions}. In
higher orders it is also necessary to consider terms with larger
degrees of $\alpha_0$. For example, in the four-loop approximation
we should also consider terms proportional to $\alpha_0^4$ in the
first equation of (\ref{NSVZ_Scheme}) and terms proportional to
$\alpha_0^2$ in the second equation of (\ref{NSVZ_Scheme}). This
allows to obtain the next coefficients $b_3$ and $g_2$. If we set
$g_1=0$, then  the NSVZ scheme corresponds to $g_2=b_1=b_2=b_3=0$.

\section{Relation between the NSVZ and $\overline{\mbox{DR}}$ schemes
in the three-loop approximation}\label{Section_DRED_NSVZ} \hspace{\parindent}

The three-loop $\beta$-function and the two-loop anomalous dimension
for an arbitrary ${\cal N}=1$ SYM theory with matter in the
$\overline{\mbox{DR}}$ scheme have been calculated in
\cite{Jack:1996vg}.\footnote{In order to obtain the results of
Ref. \cite{Jack:1996vg}, it is necessary to set $\alpha =
g^2/4\pi$, $\gamma(\alpha) = 2\gamma(g)$, $\beta(\alpha) = g
\beta(g)/2\pi$.} The result does not satisfy the NSVZ relation.
However, in \cite{Jack:1996vg} the authors obtained a special
redefinition of the coupling constant after which the NSVZ
relation is satisfied. For ${\cal N}=1$ SQED, considered here,
this redefinition is written as

\begin{equation}\label{Jones_Result}
\alpha\equiv \alpha_{\mbox{\scriptsize NSVZ}} \to \alpha'\equiv
\alpha_{\overline{\mbox{\scriptsize{DR}}}} = \alpha -
\frac{\alpha^3}{4\pi^2} + O(\alpha^4).
\end{equation}

In general, under a finite renormalization

\begin{equation}
\alpha \to \alpha'(\alpha);\qquad Z'(\alpha',\Lambda/\mu) =
z(\alpha) Z(\alpha,\Lambda/\mu)
\end{equation}

\noindent the $\beta$-function (\ref{Beta_Definition2}) and the
anomalous dimension (\ref{Gamma_Definition2}) are changed
according to the following rules:

\begin{eqnarray}\label{Beta_Transformation}
&& \widetilde \beta'(\alpha') =
\frac{d\alpha'}{d\ln\mu}\Big|_{\alpha_0=\mbox{\scriptsize const}}
= \frac{d\alpha'}{d\alpha} \widetilde
\beta(\alpha);\\
\label{Gamma_Transformation} && \widetilde \gamma'(\alpha') =
\frac{d\ln Z'}{d\ln\mu}\Big|_{\alpha_0=\mbox{\scriptsize const}} =
\frac{d\ln z}{d\alpha}\cdot \widetilde \beta(\alpha) +
\widetilde\gamma(\alpha).
\end{eqnarray}

\noindent Using these equations it is easy to see that if
$\widetilde\beta(\alpha)$ and $\widetilde\gamma(\alpha)$ satisfy
the NSVZ relation, then

\begin{equation}
\widetilde\beta'(\alpha') = \frac{d\alpha'}{d\alpha}\cdot
\frac{\alpha^2}{\pi}
\frac{1-\widetilde\gamma'(\alpha')}{1-\alpha^2 (d\ln z/d
\alpha)/\pi}\,\Big|_{\alpha=\alpha(\alpha')}.
\end{equation}

>From Eq. (\ref{Beta_Transformation}) we see that a shift
$\alpha\to \alpha+\delta\alpha$ leads to the following change of
the $\beta$-function:

\begin{equation}
\delta\widetilde\beta(\alpha) =
\widetilde\beta(\alpha)\frac{\partial
\delta\alpha}{\partial\alpha} - \delta\alpha
\frac{\partial\widetilde\beta}{\partial\alpha}.
\end{equation}

\noindent Substituting the explicit expression for $\delta\alpha$
from Eq. (\ref{Jones_Result}) we reproduce the result of
\cite{Jack:1996vg} for ${\cal N}=1$ SQED:

\begin{equation}
\delta\widetilde\beta(\alpha) = \widetilde
\beta_{\overline{\mbox{\scriptsize{DR}}}} - \widetilde
\beta_{\mbox{\scriptsize NSVZ}} = -\frac{\alpha^2}{\pi}\cdot
\frac{3\alpha^2}{4\pi^2} + \frac{\alpha^3}{4\pi^2} \cdot
\frac{2\alpha}{\pi} +O(\alpha^5) = -\frac{\alpha^4}{4\pi^3} +
O(\alpha^5).
\end{equation}

In this paper we relate the results obtained with the higher
derivative regularization with the results obtained in
\cite{Jack:1996vg}. More exactly, we construct a finite
renormalization relating the NSVZ scheme with the
$\overline{\mbox{DR}}$ scheme and boundary conditions for which
the functions $\widetilde\beta(\alpha)$ and
$\widetilde\gamma(\alpha)$, depending on the renormalized coupling
constant, coincide with the results obtained in the
$\overline{\mbox{DR}}$ scheme.

The coefficient $b_1$ is found by comparing the two-loop expression
for the anomalous dimension of the matter superfield

\begin{equation}\label{Gamma_DRED}
\widetilde\gamma_{\overline{\mbox{\scriptsize{DR}}}}(\alpha) =
-\frac{\alpha}{\pi} + \frac{\alpha^2}{\pi^2} + O(\alpha^3)
\end{equation}

\noindent with Eq. (\ref{Gamma_Answer2}). Two these expressions
coincide, if

\begin{equation}
b_1 - g_1 = \sum\limits_{I=1}^n c_I \ln a_I + \frac{1}{2}.
\end{equation}

\noindent The next coefficient $b_2$ can be found by comparing the
results for the three-loop $\beta$-function. The result of
\cite{Jack:1996vg} in our notation can be written as

\begin{equation}\label{Beta_DRED}
\widetilde\beta_{\overline{\mbox{\scriptsize{DR}}}}(\alpha) =
\frac{\alpha^2}{\pi} + \frac{\alpha^3}{\pi^2} -
\frac{5\alpha^4}{4\pi^3} + O(\alpha^5).
\end{equation}

\noindent Comparing this equation with (\ref{Beta_Answer2}) we
obtain

\begin{equation}
b_2 - g_1 = \frac{1}{4}.
\end{equation}

\noindent Therefore, for $x_0 = \ln\Lambda/\mu_0 =- g_1$

\begin{eqnarray}
&& Z_{\overline{\mbox{\scriptsize{DR}}}}(\alpha_0,x_0)=1 + O(\alpha_0^2); \nonumber\\
&&
\frac{1}{\alpha_{\overline{\mbox{\scriptsize{DR}}}}(\alpha_0,x_0)}
= \frac{1}{\alpha_0} + \frac{1}{\pi}\Big(\sum\limits_{I=1}^n c_I
\ln a_I + \frac{1}{2}\Big) + \frac{\alpha_0}{4\pi^2} +
O(\alpha_0^2).
\end{eqnarray}

\noindent In the considered approximation these equations define a
scheme in which the RG functions (\ref{Beta_Definition2}) and
(\ref{Gamma_Definition2}) coincide with the corresponding RG
functions obtained with the $\overline{\mbox{DR}}$ scheme. (It is
expedient to compare these conditions with Eq. (\ref{NSVZ_Scheme})
defining the NSVZ scheme.)

It is easy to see that $Z_{\overline{\mbox{\scriptsize{DR}}}}$ and
$\alpha_{\overline{\mbox{\scriptsize{DR}}}}$ can be obtained from
$Z_{\mbox{\scriptsize NSVZ}}$ and $\alpha_{\mbox{\scriptsize
NSVZ}}$ (constructed with the higher derivative regularization) by
the following finite renormalization:

\begin{eqnarray}\label{NSVZ_DRED}
&&\hspace*{-7mm} Z_{\overline{\mbox{\scriptsize{DR}}}}
(\alpha_{\overline{\mbox{\scriptsize{DR}}}},\Lambda/\mu)=\Big(1 -
\frac{\alpha_{\mbox{\scriptsize NSVZ}}}{\pi}
\Big(\sum\limits_{I=1}^n c_I \ln a_I + \frac{1}{2}-b_1\Big) +
O(\alpha_{\mbox{\scriptsize NSVZ}}^2)\Big) Z_{\mbox{\scriptsize
NSVZ}}(\alpha_{\mbox{\scriptsize
NSVZ}},\Lambda/\mu); \nonumber\\
&&\hspace*{-7mm}
\frac{1}{\alpha_{\overline{\mbox{\scriptsize{DR}}}}} =
\frac{1}{\alpha_{\mbox{\scriptsize NSVZ}}} + \frac{b_1}{\pi} +
\frac{\alpha_{\mbox{\scriptsize NSVZ}}}{4\pi^2} -
\frac{\alpha_{\mbox{\scriptsize
NSVZ}}}{\pi^2}\Big(\sum\limits_{I=1}^n c_I \ln a_I + \frac{1}{2} -
b_1\Big) + O(\alpha_{\mbox{\scriptsize NSVZ}}^2),
\end{eqnarray}

\noindent where $b_1$ is an arbitrary finite constant. For

\begin{equation}
b_1= \sum\limits_{I=1}^n c_I \ln a_I + \frac{1}{2}
\end{equation}

\noindent Eq. (\ref{NSVZ_DRED}) corresponds to the result of Ref.
\cite{Jack:1996vg}, which is given by Eq. (\ref{Jones_Result}),
because in \cite{Jack:1996vg} the authors construct the NSVZ
scheme in which the anomalous dimension is fixed by the condition

\begin{equation}
Z_{\overline{\mbox{\scriptsize{DR}}}}
(\alpha_{\overline{\mbox{\scriptsize{DR}}}},\Lambda/\mu)=
Z_{\mbox{\scriptsize NSVZ}}(\alpha_{\mbox{\scriptsize
NSVZ}},\Lambda/\mu).
\end{equation}

\noindent In the NSVZ scheme obtained in this paper at the
one-loop level this condition can be satisfied by a special choice
of the constant $b_1$. Fixing this additional constant we
construct a finite renormalization after which not only the
$\beta$-functions, but also the anomalous dimensions coincide. In
the next orders the situation is the same: it is possible to
redefine the coupling constant so that the NSVZ relation is valid
for the RG functions defined in terms of the renormalized coupling
constant. The finite renormalization of the matter superfields is
not fixed by this requirement. This renormalization can be found
requiring coincidence of the anomalous dimensions.

\section{Conclusion}
\hspace{\parindent}

For ${\cal N}=1$ SQED regularized by higher derivatives the NSVZ
relation is automatically satisfied, if a $\beta$-function and an
anomalous dimension are defined in terms of the bare coupling
constant according to Eqs. (\ref{Beta_Definition1}) and
(\ref{Gamma_Definition1}). If these definitions are used, then the
regularization described in Sect. \ref{Section_Regularization}
always gives the exact NSVZ $\beta$-function. In order to obtain a
result different from the exact NSVZ $\beta$-function in this
case, it is necessary to make a finite rescaling of the
Pauli--Villars masses and a  finite tuning of the bare charge
$\alpha_0$. In particular, using this procedure it is possible to
make coefficients of the RG functions defined in terms of the bare
charge equal to the ones of the RG functions defined in terms of
the renormalized charge in the $\overline{\mbox{DR}}$ scheme.

If a $\beta$-function and an anomalous dimension are defined in
terms of the renormalized coupling constant, the NSVZ
$\beta$-function is obtained only in a special (NSVZ) scheme,
which is related with the MOM scheme by a finite renormalization.
In case of using the DRED regularization the only way to construct
this finite renormalization is to use the definition of the NSVZ
scheme. However, if the theory is regularized by higher derivatives
there is a concrete extra prescription: there should be a point $x_0$
in which

\begin{equation}\label{NSVZ_Scheme_In_Conclusion}
Z_3(\alpha,x_0) = 1;\qquad Z(\alpha,x_0)=1.
\end{equation}

\noindent In order to obtain a result different from the NSVZ
$\beta$-function, it is possible to make a finite renormalization,
which, in turn, changes the boundary conditions.

These results are verified by an explicit three-loop calculation.
In particular, we relate the $\overline{\mbox{DR}}$ scheme and the
NSVZ scheme obtained with the higher derivative regularization by
a finite renormalization of both the coupling constant and the
matter superfields similar to the result of \cite{Jack:1996vg}.
However, an additional finite constant which gives the required
value of the two-loop anomalous dimension should be fixed.

\bigskip
\bigskip

\noindent {\Large\bf Acknowledgements.}

\bigskip

\noindent We are grateful to V.A.Rubakov, I.V.Tyutin, and
V.N.Velizhanin for valuable discussions.  The  work of AK
is   supported  in part by the  Grant NSh-5590.2012.2,
Russian Foundation of Basic Research grants No. 11-01-00182 and
No. 11-02-00112 and is done within the framework of the Ministry
of Science and Education contract No. 8412. The work of KS  was
supported by Russian Foundation for Basic Research grant No.
11-01-00296.

\bigskip

\appendix

\noindent {\bf \Large Appendixes}

\section{Three-loop renormalization}
\hspace{\parindent}\label{Appendix_Renormalization}

In this appendix we describe the renormalization of ${\cal N}=1$
SQED, regularized by higher derivatives, at the three-loop level
following Refs. \cite{Soloshenko:2003sx,Soloshenko:2003nc}. We
start with the expression (\ref{G(alpha0)}) for the two-point
Green function of the matter superfields. We split the bare
coupling constant $\alpha_0 = e_0^2/4\pi$ into the renormalized
coupling constant $\alpha$ and a counterterm. In the lowest
approximation this can be done according to Eq.
(\ref{One-Loop_Coupling_Renormalization}), in which the logarithm
compensate a one-loop divergence, and a finite parameter $b_1$ is
not fixed. The expression (\ref{G(alpha0)}) was calculated in Ref.
\cite{Soloshenko:2003sx} for $R_k = 1+k^{2n}/\Lambda^{2n}$. The
result can be written in the following form:

\begin{eqnarray}
&& G(\alpha_0,\Lambda/p) = 1 - \frac{\alpha_0}{\pi}
\ln\frac{\Lambda}{p} -\frac{\alpha_0}{2\pi} +
\frac{\alpha_0^2}{\pi^2} \ln{}^2 \frac{\Lambda}{p} +
\frac{\alpha_0^2}{\pi^2}\ln\frac{\Lambda}{p}\Big(\sum\limits_{I=1}^n
c_I \ln a_I +\frac{5}{2}\Big) + \frac{\alpha_0^2}{\pi^2}
c_2\nonumber\\
&& + \Big(\mbox{terms vanishing in the limit $\Lambda\to
\infty$}\Big) + O(\alpha_0^3),
\end{eqnarray}

\noindent where the constant $c_2$ was not found in Ref.
\cite{Soloshenko:2003sx}. (A finite part of this function in the
one-loop approximation has been explicitly calculated.) The finite
constants $a_I$ are related with the Pauli--Villars masses by Eq.
(\ref{Scheme}). In terms of the renormalized coupling constant the
function $G$ is given by

\begin{eqnarray}\label{G(alpha)}
&& G(\alpha,\Lambda/\mu,\Lambda/p) = 1 - \frac{\alpha}{\pi}
\Big(\ln \frac{\Lambda}{p} + \frac{1}{2}\Big)
+\frac{\alpha^2}{\pi^2} \Big[\ln{}^2 \frac{\Lambda}{p}
-\ln\frac{\Lambda}{\mu} \ln\frac{\Lambda}{p} - b_1 \ln
\frac{\Lambda}{p} - \frac{1}{2} \ln \frac{\Lambda}{\mu} -
\frac{1}{2} b_1\qquad\nonumber\\
&& + \ln\frac{\Lambda}{p} \Big(\sum\limits_{I=1}^n c_I \ln a_I
+\frac{5}{2}\Big) + c_2\Big] + (\mbox{terms vanishing in the limit
$\Lambda\to \infty$}) + O(\alpha^3).
\end{eqnarray}

\noindent This expression should be multiplied by a factor $Z$
such that the expression $ZG$ is finite. This is true, if the
renomalization constant $Z$ is given by Eq. (\ref{Two_Loop_Z}):

$$
Z = 1 + \frac{\alpha}{\pi}\Big(\ln\frac{\Lambda}{\mu}+g_1\Big)
+\frac{\alpha^2}{\pi^2}\ln^2\frac{\Lambda}{\mu}
-\frac{\alpha^2}{\pi^2}\ln\frac{\Lambda}{\mu}\Big(\sum\limits_{I=1}^n
c_I\ln a_I - b_1 + \frac{3}{2} - g_1\Big) + \frac{\alpha^2}{\pi^2}
g_2 + O(\alpha^3),
$$

\noindent where $g_1$ and $g_2$ are undefined finite constants.
Really, calculating the product $Z G$ in the limit $\Lambda\to \infty$,
it is easy to see that all terms containing $\ln \Lambda$ cancel
each other and

\begin{eqnarray}
&& G_{\mbox{\scriptsize ren}}(\alpha,\mu/p) =
\lim\limits_{\Lambda\to \infty} ZG = 1 - \frac{\alpha}{\pi}
\Big(\ln \frac{\mu}{p} - g_1 +\frac{1}{2}\Big) +
\frac{\alpha^2}{\pi^2} \Big[\ln{}^2 \frac{\mu}{p} +
\ln\frac{\mu}{p} \Big(\sum\limits_{I=1}^n
c_I \ln a_I +\frac{5}{2}\Big)\nonumber\\
&& - (b_1+g_1) \ln\frac{\mu}{p} + c_2 - \frac{1}{2} g_1  + g_2 -
\frac{1}{2} b_1\Big] + O(\alpha^3).\qquad
\end{eqnarray}

\noindent This expression is finite for all finite values of
$b_1$, $g_1$, and $g_2$. Therefore, they can be chosen in an
arbitrary way (unlike the constant $c_2$, which is fixed, but
unknown).

Values of the constants $b_i$ and $g_i$ can be found from boundary
conditions, which specify the scheme. For example, the MOM scheme is
defined by the boundary conditions
(\ref{MOM_Boundary_Conditions}):

$$
Z_{\mbox{\scriptsize MOM}}G(\alpha_{\mbox{\scriptsize MOM}},p=\mu)
= 1; \qquad d^{-1}(\alpha_{\mbox{\scriptsize
MOM}},p=\mu)=\alpha_{\mbox{\scriptsize MOM}}^{-1}.
$$

\noindent In the considered case the first condition gives

\begin{equation}
g_1 = \frac{1}{2};\qquad g_2 = - c_2 + \frac{1}{2} g_1
+\frac{1}{2} b_1 = - c_2 + \frac{1}{4} +\frac{1}{2} b_1.
\end{equation}

\noindent (The constant $b_1$ is defined from the other boundary
condition.) Therefore, in the subtraction scheme defined by the
boundary conditions (\ref{MOM_Boundary_Conditions}) the constants
$b_i$ and $g_i$ are related with the finite parts of the Green
function ($c_1 = -1/2$ and $c_2$).

However, it is possible to impose the different boundary
conditions (\ref{NSVZ_Scheme_Z3}):

$$
Z_{\mbox{\scriptsize NSVZ}}(\alpha_{\mbox{\scriptsize
NSVZ}},\mu=\Lambda)=1; \qquad (Z_3)_{\mbox{\scriptsize
NSVZ}}(\alpha_{\mbox{\scriptsize NSVZ}},\mu=\Lambda)=1.
$$

\noindent In this case

\begin{equation}
1 = Z_{\mbox{\scriptsize NSVZ}}(\alpha_{\mbox{\scriptsize
NSVZ}},\Lambda=\mu) = 1 + \frac{\alpha_{\mbox{\scriptsize
NSVZ}}}{\pi}g_1 + \frac{\alpha^2_{\mbox{\scriptsize NSVZ}}}{\pi^2}
g_2 + O(\alpha_{\mbox{\scriptsize NSVZ}}^3),
\end{equation}

\noindent so that

\begin{equation}
g_1=0;\qquad g_2=0.
\end{equation}

\noindent In this case the constants $g_i$ (and $b_i$, see below)
are independent of the finite parts of the effective action (i.e.
the constants $-1/2$ and $c_2$).

Let us proceed to the renormalization of the coupling constant.
For this purpose we should write the function $d^{-1}$. The
integrals defining this function in the three-loop approximation
have been found in Ref. \cite{Soloshenko:2003nc}. Although it is
very difficult to calculate their finite parts, we know
\cite{Stepanyantz:2011jy} that this function satisfies Eq.
(\ref{Green_Function_Relation}):

$$
\frac{d}{d\ln\Lambda} \Big(d^{-1}(\alpha_0,\Lambda/p) -
\alpha_0^{-1}\Big)\Big|_{p=0}  =
\frac{1}{\pi}\Big(1-\frac{d}{d\ln\Lambda} \ln
G(\alpha_0,\Lambda/q)\Big|_{q=0}\Big).
$$

\noindent This equation gives the NSVZ relation in terms of the bare
charge, i.e. for the RG functions (\ref{Beta_Definition1}) and
(\ref{Gamma_Definition1}). The first term in the left hand side
vanishes, because the function $d^{-1}$ depends only on $\alpha$
and $\mu/p$ and (expressed in terms of the renormalized coupling
constant) is independent of $\Lambda$ in the limit $p\to 0$. The
right hand side of this equation can be calculated differentiating
the logarithm of Eq. (\ref{G(alpha)}), taking into account that
the limit $\Lambda\to \infty$ in the massless case is equivalent
to the limit $q\to 0$ due to the dependence on $q/\Lambda$. Then it is
easy to see that all terms containing the dependence on the
momentum $q$ vanish. Expressing the result in terms of the bare
coupling constant we obtain Eq. (\ref{Beta_Expression}):

$$
- \frac{d}{d\ln\Lambda}(\alpha_0^{-1}) = \frac{1}{\pi} +
\frac{\alpha_0}{\pi^2}  - \frac{\alpha_0^2}{\pi^3}
\Big(\sum\limits_{I=1}^n c_I \ln a_I +\frac{3}{2}\Big) +
O(\alpha_0^3).
$$

\noindent A renormalization of the coupling constant can be found
by integrating this equation, which should be written in terms of
the renormalized coupling constant:

\begin{equation}
- \frac{d}{d\ln\Lambda} (\alpha_0^{-1}) = \frac{1}{\pi} +
\frac{\alpha}{\pi^2} +\frac{\alpha^2}{\pi^3} \ln
\frac{\Lambda}{\mu} +\frac{\alpha^2}{\pi^3} b_1 -
\frac{\alpha^2}{\pi^3} \Big(\sum\limits_{I=1}^n c_I \ln a_I
+\frac{3}{2}\Big) + O(\alpha^3).
\end{equation}

\noindent The result is given by Eq. (\ref{Alpha0}):

\begin{eqnarray}
&& \frac{1}{\alpha_0} = \frac{1}{\alpha} - \frac{1}{\pi} \Big(
\ln\frac{\Lambda}{\mu} + b_1\Big) - \frac{\alpha}{\pi^2}
\Big(\ln\frac{\Lambda}{\mu} + b_2\Big)\nonumber\\
&& \qquad\qquad\qquad -
\frac{\alpha^2}{\pi^3}\Big(\frac{1}{2}\ln^2 \frac{\Lambda}{\mu}
-\ln \frac{\Lambda}{\mu} \sum\limits_{I=1}^n c_I \ln a_I
-\frac{3}{2} \ln \frac{\Lambda}{\mu} + b_1 \ln\frac{\Lambda}{\mu}
+ b_3\Big) + O(\alpha^3),\qquad\nonumber
\end{eqnarray}

\noindent where $b_1$, $b_2$, and $b_3$ are integration constants
in each order.

Although we have not calculated the inverse invariant charge
$d^{-1}$, starting from this equation it is possible to restore
its divergent part by requiring its finiteness after the
substitution $\alpha_0 \to \alpha_0(\alpha,\Lambda/\mu)$:

\begin{eqnarray}
&& d^{-1}(\alpha_0,\Lambda/p) = \frac{1}{\alpha_0} +
\frac{1}{\pi}\Big(\ln\frac{\Lambda}{p} + d_1\Big) +
\frac{\alpha_0}{\pi^2}\Big(\ln\frac{\Lambda}{p} +
d_2\Big)\nonumber\\
&& + \frac{\alpha_0^2}{\pi^3}\Big(-\frac{1}{2}\ln^2
\frac{\Lambda}{p} -\ln \frac{\Lambda}{p} \sum\limits_{I=1}^n c_I
\ln a_I -\frac{3}{2} \ln \frac{\Lambda}{p} - d_2
\ln\frac{\Lambda}{p} + d_3\Big)\nonumber\\
&& + \mbox{(terms vanishing in the limit $\Lambda\to \infty$)} +
O(\alpha_0^3),\vphantom{\Big(}
\end{eqnarray}

\noindent where $d_1$, $d_2$, and $d_3$ are finite constants,
which should be found by calculating Feynman graphs. It is
easy to see that for any values of $d_i$ and for an arbitrary
choice of $b_i$ the above expression for $d^{-1}$ is a finite
function of $\alpha$ and $\mu/p$:

\begin{eqnarray}
&& d^{-1}\Big(\alpha_0(\alpha,\Lambda/\mu),\Lambda/p\Big) =
\frac{1}{\alpha} + \frac{1}{\pi}\Big(\ln\frac{\mu}{p} + d_1 -
b_1\Big) + \frac{\alpha}{\pi^2}\Big(\ln\frac{\mu}{p} + d_2 -
b_2\Big)\nonumber\\
&& + \frac{\alpha^2}{\pi^3}\Big(-\frac{1}{2}\ln^2 \frac{\mu}{p}
-\ln \frac{\mu}{p} \sum\limits_{I=1}^n c_I \ln a_I -\frac{3}{2}
\ln \frac{\mu}{p} + (b_1 - d_2) \ln\frac{\mu}{p} + d_3 - b_3 + b_1
d_2\Big)\nonumber\\
&& + \mbox{(terms vanishing in the limit $\Lambda\to \infty$)} +
O(\alpha^3).\vphantom{\Big(}
\end{eqnarray}

The coefficients $b_i$ can be found from boundary conditions. If
the boundary conditions (\ref{MOM_Boundary_Conditions}), defining
the MOM scheme, are imposed, we obtain:

\begin{equation}
b_1 = d_1;\qquad b_2 = d_2;\qquad b_3 = d_3 + b_1 d_2 = d_3+ d_1
d_2.
\end{equation}

\noindent Therefore, in this case the coefficients defining the
scheme are expressed in terms of the finite parts of the Green
function. In the limit $\Lambda\to \infty$ it is easy to see that
in the MOM scheme

\begin{eqnarray}
&& Z_{\mbox{\scriptsize MOM}}G(\alpha_{\mbox{\scriptsize
MOM}},\mu/p)
= 1 - \frac{\alpha_{\mbox{\scriptsize MOM}}}{\pi} \ln \frac{\mu}{p}\\
&&\qquad\qquad\qquad\qquad\qquad + \frac{\alpha_{\mbox{\scriptsize
MOM}}^2}{\pi^2} \Big(\ln{}^2 \frac{\mu}{p} + \ln \frac{\mu}{p}
\Big( \sum\limits_{I=1}^n c_I \ln a_I + 2 -d_1\Big) \Big)
+ O(\alpha_{\mbox{\scriptsize MOM}}^3);\nonumber\\
&& d^{-1}(\alpha_{\mbox{\scriptsize MOM}},\mu/p) =
\frac{1}{\alpha_{\mbox{\scriptsize MOM}}} + \frac{1}{\pi}
\ln\frac{\mu}{p} + \frac{\alpha_{\mbox{\scriptsize MOM}}}{\pi^2} \ln\frac{\mu}{p}\\
&&\qquad\ + \frac{\alpha_{\mbox{\scriptsize
MOM}}^2}{\pi^3}\Big(-\frac{1}{2}\ln^2 \frac{\mu}{p} -\ln
\frac{\mu}{p} \sum\limits_{I=1}^n c_I \ln a_I -\frac{3}{2} \ln
\frac{\mu}{p} + (d_1 - d_2) \ln\frac{\mu}{p}\Big) +
O(\alpha_{\mbox{\scriptsize MOM}}^3).\qquad\nonumber
\end{eqnarray}

If we impose the boundary conditions (\ref{NSVZ_Scheme_Z3}),
defining the NSVZ scheme, then the coefficients $b_i$ do not
depend on the finite parts of the Green function:

\begin{equation}
\frac{1}{\alpha_{\mbox{\scriptsize NSVZ}}} =  \frac{1}{\alpha_0} =
\frac{1}{\alpha_{\mbox{\scriptsize NSVZ}}} - \frac{1}{\pi} b_1 -
\frac{\alpha_{\mbox{\scriptsize NSVZ}}}{\pi^2} b_2 -
\frac{\alpha_{\mbox{\scriptsize NSVZ}}^2}{\pi^3} b_3 +
O(\alpha_{\mbox{\scriptsize NSVZ}}^3),
\end{equation}

\noindent so that

\begin{equation}
b_1=0; \qquad b_2=0; \qquad b_3=0.
\end{equation}

\noindent In this case the renormalized Green functions in the
limit $\Lambda\to \infty$ are written as

\begin{eqnarray}
&& Z_{\mbox{\scriptsize NSVZ}}G(\alpha_{\mbox{\scriptsize
NSVZ}},\mu/p) = 1 - \frac{\alpha_{\mbox{\scriptsize NSVZ}}}{\pi}
\Big(\ln \frac{\mu}{p}+\frac{1}{2}\Big)\\
&&\qquad\qquad\qquad\qquad\quad + \frac{\alpha_{\mbox{\scriptsize
NSVZ}}^2}{\pi^2}\Big[\ln{}^2 \frac{\mu}{p} + \ln\frac{\mu}{p}
\Big(\sum\limits_{I=1}^n c_I \ln a_I + \frac{5}{2}\Big) +
c_2\Big] + O(\alpha_{\mbox{\scriptsize NSVZ}}^3);\nonumber\\
&& d^{-1}(\alpha_{\mbox{\scriptsize NSVZ}},\mu/p) =
\frac{1}{\alpha_{\mbox{\scriptsize NSVZ}}} +
\frac{1}{\pi}\Big(\ln\frac{\mu}{p} + d_1\Big) +
\frac{\alpha_{\mbox{\scriptsize NSVZ}}}{\pi^2}\Big(\ln\frac{\mu}{p} + d_2\Big)\\
&&\qquad + \frac{\alpha_{\mbox{\scriptsize
NSVZ}}^2}{\pi^3}\Big(-\frac{1}{2}\ln^2 \frac{\mu}{p} -\ln
\frac{\mu}{p} \sum\limits_{I=1}^n c_I \ln a_I -\frac{3}{2} \ln
\frac{\mu}{p} - d_2 \ln\frac{\mu}{p} + d_3\Big) +
O(\alpha_{\mbox{\scriptsize NSVZ}}^3).\qquad\nonumber
\end{eqnarray}

It is easy to verify that the MOM scheme defined by the boundary
conditions (\ref{MOM_Boundary_Conditions}) and the NSVZ scheme
defined by the boundary conditions (\ref{NSVZ_Scheme_Z3}) are
related by the following finite renormalization:

\begin{eqnarray}
&& \frac{1}{\alpha_{\mbox{\scriptsize NSVZ}}} =
\frac{z_3(\alpha_{\mbox{\scriptsize
MOM}})}{\alpha_{\mbox{\scriptsize MOM}}} =
\frac{1}{\alpha_{\mbox{\scriptsize MOM}}}- \frac{1}{\pi} d_1 -
\frac{\alpha_{\mbox{\scriptsize MOM}}}{\pi^2} d_2 -
\frac{\alpha_{\mbox{\scriptsize MOM}}^2}{\pi^3} (d_3+ d_1
d_2) + O(\alpha_{\mbox{\scriptsize MOM}}^3);\nonumber\\
&& Z_{\mbox{\scriptsize NSVZ}}(\alpha_{\mbox{\scriptsize
NSVZ}},\Lambda/\mu) = z(\alpha_{\mbox{\scriptsize MOM}})
Z_{\mbox{\scriptsize MOM}}(\alpha_{\mbox{\scriptsize
MOM}},\Lambda/\mu)
\vphantom{\frac{1}{2}}\\
&&\qquad\qquad\qquad = \Big(1-\frac{\alpha_{\mbox{\scriptsize
MOM}}}{2\pi} - \frac{\alpha_{\mbox{\scriptsize
MOM}}^2}{\pi^2}\Big(\frac{1}{2} d_1-c_2\Big) +
O(\alpha_{\mbox{\scriptsize MOM}}^3)\Big) Z_{\mbox{\scriptsize
MOM}}(\alpha_{\mbox{\scriptsize MOM}},\Lambda/\mu),
\nonumber\qquad
\end{eqnarray}

\noindent where the remaining coefficients can be found by
calculating finite parts of the Feynman diagrams.

\section{Existence of the NSVZ scheme}
\hspace{\parindent}\label{Appendix_Existence}

In this section we prove that the scheme defined by Eq.
(\ref{NSVZ_Scheme}) exists and without loss of generality it is
possible to set $x_0=0$. For this purpose we note that the
renormalized coupling constant $\alpha$ is defined so that the
function $d^{-1}\Big(\alpha_0(\alpha,\Lambda/\mu),\Lambda/p\Big)$
does not depend on $\Lambda$ in the limit $\Lambda\to \infty$.
Certainly, this condition does not uniquely define $\alpha$. In
particular, it is possible to perform a finite redefinition of the
coupling constant $\alpha = \alpha(\alpha')$. After this
redefinition

\begin{equation}
\alpha_0(\alpha,\Lambda/\mu) \to
\alpha_0(\alpha(\alpha'),\Lambda/\mu).
\end{equation}

\noindent Because the function $\alpha(\alpha')$ is finite, the
expression

\begin{equation}
d^{-1}\Big(\alpha_0(\alpha(\alpha'),\Lambda/\mu),\Lambda/p\Big)
\end{equation}

\noindent is also finite. Therefore, $\alpha'$ can be also chosen
as a renormalized coupling constant.

Let us choose the renormalized coupling constant in an arbitrary
way. It is defined by a function $\alpha_0(\alpha,\Lambda/\mu)$.
Then we construct a new function

\begin{equation}\label{B_Definition}
\alpha_0(\alpha,x=0) \equiv b(\alpha),
\end{equation}

\noindent where

\begin{equation}
x \equiv \ln\frac{\Lambda}{\mu}.
\end{equation}

\noindent Evidently, the function $b(\alpha)$ does not depend on
$\Lambda$ and is finite in the limit $\Lambda\to \infty$. Then we
define a new renormalized coupling constant according to the
prescription

\begin{equation}
\alpha'(\alpha) \equiv b(\alpha).
\end{equation}

\noindent Then, from Eq. (\ref{B_Definition}) we obtain

\begin{equation}
\alpha_0(\alpha(\alpha'),x=0) = \alpha'.
\end{equation}

\noindent Therefore, choosing $\alpha'$ as a renormalized coupling
constant we satisfy the first condition in Eq. (\ref{NSVZ_Scheme})
with $x_0=0$.

Similarly, by definition, the renormalization constant $Z$ is
chosen in such a way that the expression

\begin{equation}
Z(\alpha,\Lambda/\mu)
G\Big(\alpha_0(\alpha,\Lambda/\mu),\Lambda/p\Big)
\end{equation}

\noindent is finite. Again, this condition does not uniquely
define the function $Z$. In particular, it is possible to multiply
$Z$ by an arbitrary finite function of the renormalized coupling
constant $\alpha$. This possibility can be used for constructing
the renormalization constant which satisfies the second condition
in Eq. (\ref{NSVZ_Scheme}). For this purpose we consider an
arbitrary renormalization constant $Z$ and construct the function

\begin{equation}\label{G_Definition}
Z(\alpha,x=0) \equiv g(\alpha).
\end{equation}

\noindent Then the renormalization constant $Z'$ is defined
according to the prescription

\begin{equation}\label{Z_Prime_Definition}
Z'(\alpha',\Lambda/\mu) \equiv \frac{1}{g(\alpha(\alpha'))}
Z\Big(\alpha(\alpha'),\Lambda/\mu\Big).
\end{equation}

\noindent It is evident that the expression

\begin{equation}
Z'(\alpha',\Lambda/\mu)
G\Big(\alpha_0(\alpha',\Lambda/\mu),\Lambda/p\Big)
\end{equation}

\noindent does not depend on $\Lambda$ in the limit $\Lambda\to
\infty$. Therefore the function $Z'(\alpha',\Lambda/\mu)$ can be
chosen as a renormalization constant. Moreover, due to Eqs.
(\ref{G_Definition}) and (\ref{Z_Prime_Definition}) this function
satisfies the condition

\begin{equation}
Z'(\alpha',x=0) = 1.
\end{equation}

\noindent Therefore, we have constructed the renormalization
constants $Z_3'(\alpha',\Lambda/\mu)$ and
$Z'(\alpha',\Lambda/\mu)$ which satisfy the conditions
(\ref{NSVZ_Scheme_Z3}) with $x_0=0$.

Let us verify these results by an explicit three-loop calculation.
>From Eq. (\ref{Alpha0}) we obtain

\begin{equation}
\frac{1}{\alpha_0(\alpha,x=0)} = \frac{1}{\alpha} - \frac{1}{\pi}
b_1 - \frac{\alpha}{\pi^2} b_2 - \frac{\alpha^2}{\pi^3} b_3 +
O(\alpha^3).
\end{equation}

\noindent As a consequence, the function $b(\alpha)$ defined by
Eq. (\ref{B_Definition}) is given by

\begin{equation}\label{Three_Loop_Alpha_Prime}
b(\alpha) = \alpha \Big(1 - \frac{\alpha}{\pi} b_1 -
\frac{\alpha^2}{\pi^2} b_2 - \frac{\alpha^3}{\pi^3} b_3\Big)^{-1}
+ O(\alpha^5)\equiv \alpha'(\alpha).
\end{equation}

\noindent Rewriting the bare coupling constant $\alpha_0$ in terms
of $\alpha'$ defined by this equation we obtain

\begin{equation}
\frac{1}{\alpha_0} = \frac{1}{\alpha'} - \frac{1}{\pi}
\ln\frac{\Lambda}{\mu} - \frac{\alpha'}{\pi^2}
\ln\frac{\Lambda}{\mu} -\frac{(\alpha')^2}{\pi^3}
\Big(\frac{1}{2}\ln{}^2 \frac{\Lambda}{\mu} -
\ln\frac{\Lambda}{\mu} \sum\limits_{I=1}^n c_I \ln
\frac{M_I}{\Lambda} - \frac{3}{2} \ln \frac{\Lambda}{\mu}\Big) +
O\left((\alpha')^3\right).
\end{equation}

\noindent This choice is a particular case of Eq. (\ref{Alpha0})
with $b_1=b_2=b_3=0$. Therefore, the function
$d^{-1}\left(\alpha_0(\alpha',\Lambda/\mu),\Lambda/p\right)$ does
not depend on $\Lambda$ in the limit $\Lambda\to \infty$.
Moreover, it is evident that

\begin{equation}\label{Three_Loop_NSVZ_Scheme_Alpha}
\alpha_0(\alpha',x=0) = \alpha' + O\left((\alpha')^5\right).
\end{equation}

Similarly, one can construct the function
$Z'(\alpha',\Lambda/\mu)$ in the two-loop approximation. As a
starting point we consider the renormalization constant
(\ref{Two_Loop_Z}). Then we find

\begin{equation}
g(\alpha)= Z(\alpha,x=0) = 1 + \frac{\alpha}{\pi} g_1 +
\frac{\alpha^2}{\pi^2} g_2 + O(\alpha^3).
\end{equation}

\noindent Writing this function in terms of $\alpha'$ defined by
Eq. (\ref{Three_Loop_Alpha_Prime}) we obtain

\begin{equation}
g\left(\alpha(\alpha')\right) = 1 + \frac{\alpha'}{\pi} g_1 +
\frac{(\alpha')^2}{\pi^2} (g_2 - g_1 b_1) +
O\left((\alpha')^3\right).
\end{equation}

\noindent As a consequence,

\begin{eqnarray}
&& Z'(\alpha',\Lambda/\mu) = g\left(\alpha(\alpha')\right)^{-1}
Z\Big(\alpha(\alpha'),\Lambda/\mu\Big)\nonumber\\
&& = 1 + \frac{\alpha'}{\pi}\ln\frac{\Lambda}{\mu} +
\frac{(\alpha')^2}{\pi^2} \ln{}^2\frac{\Lambda}{\mu} -
\frac{(\alpha')^2}{\pi^2}\Big( \sum\limits_{I=1}^n c_I \ln
\frac{M_I}{\Lambda} + \frac{3}{2}\Big) +
O\left((\alpha')^3\right).
\end{eqnarray}

\noindent This function can be obtained from Eq.
(\ref{Two_Loop_Z}) if one sets $g_1=g_2 = 0$. (Earlier we have
already set $b_1=0$.) Therefore, the finiteness of the renormalized
two-point Green function of the matter superfields is guaranteed.
Moreover, the function $Z'$ evidently satisfies the required
condition

\begin{equation}\label{Two_Loop_NSVZ_Scheme_Z}
Z'(\alpha',x=0) = 1 + O\left((\alpha')^3\right).
\end{equation}

\noindent Thus, the NSVZ scheme is constructed in the three-loop
approximation. Certainly, this procedure can be generalized to
higher orders. For example, considering the next order of the
perturbation theory we find a scheme in which Eq.
(\ref{Three_Loop_NSVZ_Scheme_Alpha}) is valid also for terms
proportional to $\alpha_0^5$, and Eq.
(\ref{Two_Loop_NSVZ_Scheme_Z}) is valid for terms proportional to
$\alpha_0^3$.

%%%%%%%%%%%%%%%%%%%%%%%%%%%%%%%%%%%%%%%%%%%%%%%%%%%%%%%%%%%%%%%%%%%%%%%%%


\begin{thebibliography}{100}

%\cite{Ogievetsky:1975nu}
\bibitem{Ogievetsky:1975nu}
  V.~I.~Ogievetsky and L.~Mezincescu,
  %``Symmetries Between Bosons and Fermions and Superfields,''
  Sov.\ Phys.\ Usp.\  {\bf 18} (1975) 960
   [Usp.\ Fiz.\ Nauk {\bf 117} (1975) 637].
  %%CITATION = SOPUA,18,960;%%

%\cite{West:1990tg}
\bibitem{West:1990tg}
  P.~C.~West,
  ``Introduction to supersymmetry and supergravity,''
  Singapore, Singapore: World Scientific (1990) 425 p.

%\cite{Buchbinder:1998qv}
\bibitem{Buchbinder:1998qv}
  I.~L.~Buchbinder and S.~M.~Kuzenko,
  ``Ideas and methods of supersymmetry and supergravity: Or a walk through superspace,''
  Bristol, UK: IOP (1998) 656 p.

%\cite{Diakonov:2010zzb}
\bibitem{Diakonov:2010zzb}
``Subtleties in quantum field theory: Lev Lipatov Festschrift,''
  Gatchina, Russia: St. Petersburg Nucl. Phys. Inst. (2010) 267 pp,
ed. D.I.Diakonov.


%\cite{Grisaru:1982zh}
\bibitem{Grisaru:1982zh}
  M.~T.~Grisaru and W.~Siegel,
  %``Supergraphity. 2. Manifestly Covariant Rules and Higher Loop Finiteness,''
  Nucl.\ Phys.\ B {\bf 201} (1982) 292
   [Erratum-ibid.\ B {\bf 206} (1982) 496].
  %%CITATION = NUPHA,B201,292;%%

%\cite{Mandelstam:1982cb}
\bibitem{Mandelstam:1982cb}
  S.~Mandelstam,
  %``Light Cone Superspace and the Ultraviolet Finiteness of the N=4 Model,''
  Nucl.\ Phys.\ B {\bf 213} (1983) 149.
  %%CITATION = NUPHA,B213,149;%%

%\cite{Brink:1982pd}
\bibitem{Brink:1982pd}
  L.~Brink, O.~Lindgren and B.~E.~W.~Nilsson,
  %``N=4 Yang-Mills Theory on the Light Cone,''
  Nucl.\ Phys.\ B {\bf 212} (1983) 401.
  %%CITATION = NUPHA,B212,401;%%



%\cite{Howe:1983sr}
\bibitem{Howe:1983sr}
  P.~S.~Howe, K.~S.~Stelle and P.~K.~Townsend,
  %``Miraculous Ultraviolet Cancellations in Supersymmetry Made Manifest,''
  Nucl.\ Phys.\ B {\bf 236} (1984) 125.
  %%CITATION = NUPHA,B236,125;%%

%\cite{Avdeev:1980bh}
\bibitem{Avdeev:1980bh}
  L.~V.~Avdeev, O.~V.~Tarasov and A.~A.~Vladimirov,
  %``Vanishing Of The Three Loop Charge Renormalization Function In A Supersymmetric Gauge Theory,''
  Phys.\ Lett.\ B {\bf 96} (1980) 94.
  %%CITATION = PHLTA,B96,94;%%

%\cite{Siegel:1979wq}
\bibitem{Siegel:1979wq}
  W.~Siegel,
  %``Supersymmetric Dimensional Regularization via Dimensional Reduction,''
  Phys.\ Lett.\ B {\bf 84} (1979) 193.
  %%CITATION = PHLTA,B84,193;%%

%\cite{Grisaru:1980nk}
\bibitem{Grisaru:1980nk}
  M.~T.~Grisaru, M.~Rocek and W.~Siegel,
  %``Zero Three Loop beta Function in N=4 Superyang-Mills Theory,''
  Phys.\ Rev.\ Lett.\  {\bf 45} (1980) 1063.
  %%CITATION = PRLTA,45,1063;%%

%\cite{Caswell:1980ru}
\bibitem{Caswell:1980ru}
  W.~E.~Caswell and D.~Zanon,
  %``Zero Three Loop Beta Function In The N=4 Supersymmetric Yang-mills Theory,''
  Nucl.\ Phys.\ B {\bf 182} (1981) 125.
  %%CITATION = NUPHA,B182,125;%%

%\cite{'tHooft:1972fi}
\bibitem{'tHooft:1972fi}
  G.~'t Hooft and M.~J.~G.~Veltman,
  %``Regularization and Renormalization of Gauge Fields,''
  Nucl.\ Phys.\ B {\bf 44} (1972) 189.
  %%CITATION = NUPHA,B44,189;%%

%\cite{Bollini:1972ui}
\bibitem{Bollini:1972ui}
  C.~G.~Bollini and J.~J.~Giambiagi,
  %``Dimensional Renormalization: The Number of Dimensions as a Regularizing Parameter,''
  Nuovo Cim.\ B {\bf 12} (1972) 20.
  %%CITATION = NUCIA,B12,20;%%

%\cite{Ashmore:1972uj}
\bibitem{Ashmore:1972uj}
  J.~F.~Ashmore,
  %``A Method of Gauge Invariant Regularization,''
  Lett.\ Nuovo Cim.\  {\bf 4} (1972) 289.
  %%CITATION = NCLTA,4,289;%%

%\cite{Cicuta:1972jf}
\bibitem{Cicuta:1972jf}
  G.~M.~Cicuta and E.~Montaldi,
  %``Analytic renormalization via continuous space dimension,''
  Lett.\ Nuovo Cim.\  {\bf 4} (1972) 329.
  %%CITATION = NCLTA,4,329;%%

%\cite{Delbourgo:1974az}
\bibitem{Delbourgo:1974az}
  R.~Delbourgo and V.~B.~Prasad,
  %``Supersymmetry in the Four-Dimensional Limit,''
  J.\ Phys.\ G {\bf 1} (1975) 377.

%\cite{Bardeen:1978yd}
\bibitem{Bardeen:1978yd}
  W.~A.~Bardeen, A.~J.~Buras, D.~W.~Duke and T.~Muta,
  %``Deep Inelastic Scattering Beyond the Leading Order in Asymptotically Free Gauge Theories,''
  Phys.\ Rev.\ D {\bf 18} (1978) 3998.
  %%CITATION = PHRVA,D18,3998;%%

%\cite{Kataev:1988sq}
\bibitem{Kataev:1988sq}
  A.~L.~Kataev, M.~D.~Vardiashvili,
  %``Scheme Dependence Of The Perturbative Series For A Physical Quantity In The G Phi**4 Theory,''
  Phys.\ Lett.\ B {\bf 221} (1989) 377
[Erratum-ibid.\ B {\bf 241} (1990) 644].
%%CITATION = PHLTA,B221,377;%%

%\cite{Siegel:1980qs}
\bibitem{Siegel:1980qs}
  W.~Siegel,
  %``Inconsistency of Supersymmetric Dimensional Regularization,''
  Phys.\ Lett.\ B {\bf 94} (1980) 37.
  %%CITATION = PHLTA,B94,37;%%

%\cite{Avdeev:1981vf}
\bibitem{Avdeev:1981vf}
  L.~V.~Avdeev, G.~A.~Chochia and A.~A.~Vladimirov,
  %``On The Scope Of Supersymmetric Dimensional Regularization,''
  Phys.\ Lett.\ B {\bf 105} (1981) 272.
  %%CITATION = PHLTA,B105,272;%%

%\cite{Avdeev:1982np}
\bibitem{Avdeev:1982np}
  L.~V.~Avdeev,
  %``Noninvariance Of Regularization By Dimensional Reduction: An Explicit Example Of Supersymmetry Breaking,''
  Phys.\ Lett.\ B {\bf 117} (1982) 317.
  %%CITATION = PHLTA,B117,317;%%

%\cite{Avdeev:1982xy}
\bibitem{Avdeev:1982xy}
  L.~V.~Avdeev and A.~A.~Vladimirov,
  %``Dimensional Regularization And Supersymmetry,''
  Nucl.\ Phys.\ B {\bf 219} (1983) 262.
  %%CITATION = NUPHA,B219,262;%%

%\cite{Velizhanin:2010vw}
\bibitem{Velizhanin:2010vw}
  V.~N.~Velizhanin,
  %``Vanishing of the four-loop charge renormalization function in N=4 SYM theory,''
  Phys.\ Lett.\ B {\bf 696} (2011) 560.
  %%CITATION = ARXIV:1008.2198;%%

%\cite{Howe:1983wj}
\bibitem{Howe:1983wj}
  P.~S.~Howe, K.~S.~Stelle and P.~C.~West,
  %``A Class of Finite Four-Dimensional Supersymmetric Field Theories,''
  Phys.\ Lett.\ B {\bf 124} (1983) 55.
  %%CITATION = PHLTA,B124,55;%%

%\cite{Avdeev:1981ew}
\bibitem{Avdeev:1981ew}
  L.~V.~Avdeev and O.~V.~Tarasov,
  %``The Three Loop Beta Function In The N=1, N=2, N=4 Supersymmetric Yang-mills Theories,''
  Phys.\ Lett.\ B {\bf 112} (1982) 356.
  %%CITATION = PHLTA,B112,356;%%

%\cite{Jack:2007ni}
\bibitem{Jack:2007ni}
  I.~Jack, D.~R.~T.~Jones, P.~Kant and L.~Mihaila,
  %``The Four-loop DRED gauge beta-function and fermion mass anomalous dimension for general gauge groups,''
  JHEP {\bf 0709} (2007) 058.
  %%CITATION = ARXIV:0707.3055;%%

%\cite{Velizhanin:2008rw}
\bibitem{Velizhanin:2008rw}
  V.~N.~Velizhanin,
  %``Three-loop renormalization of the N=1, N=2, N=4 supersymmetric Yang-Mills theories,''
  Nucl.\ Phys.\ B {\bf 818} (2009) 95.
  %%CITATION = ARXIV:0809.2509;%%

%\cite{Grisaru:1979wc}
\bibitem{Grisaru:1979wc}
  M.~T.~Grisaru, W.~Siegel and M.~Rocek,
  %``Improved Methods for Supergraphs,''
  Nucl.\ Phys.\ B {\bf 159} (1979) 429.
  %%CITATION = NUPHA,B159,429;%%

%\cite{Novikov:1983uc}
\bibitem{Novikov:1983uc}
  V.~A.~Novikov, M.~A.~Shifman, A.~I.~Vainshtein and V.~I.~Zakharov,
  %``Exact Gell-Mann-Low Function of Supersymmetric Yang-Mills Theories from Instanton Calculus,''
  Nucl.\ Phys.\ B {\bf 229} (1983) 381.
  %%CITATION = NUPHA,B229,381;%%

%\cite{Novikov:1985rd}
\bibitem{Novikov:1985rd}
  V.~A.~Novikov, M.~A.~Shifman, A.~I.~Vainshtein and V.~I.~Zakharov,
  %``Beta Function in Supersymmetric Gauge Theories: Instantons Versus Traditional Approach,''
  Phys.\ Lett.\ B {\bf 166} (1986) 329; Sov.\ J.\ Nucl.\ Phys.\  {\bf 43}
(1986) 294; [Yad.\ Fiz.\  {\bf 43} (1986) 459.]
  %%CITATION = PHLTA,B166,329;%%

%\cite{Shifman:1986zi}
\bibitem{Shifman:1986zi}
  M.~A.~Shifman and A.~I.~Vainshtein,
  %``Solution of the Anomaly Puzzle in SUSY Gauge Theories and the Wilson Operator Expansion,''
  Nucl.\ Phys.\ B {\bf 277}  (1986) 456; Sov.\ Phys.\ JETP {\bf 64} (1986) 428;
[Zh.\ Eksp.\ Teor.\ Fiz.\  {\bf 91}  (1986) 723.]
  %%CITATION = NUPHA,B277,456;%%

%\cite{Vainshtein:1986ja}
\bibitem{Vainshtein:1986ja}
  A.~I.~Vainshtein, V.~I.~Zakharov and M.~A.~Shifman,
  %``Gell-mann-low Function In Supersymmetric Electrodynamics,''
  JETP Lett.\  {\bf 42} (1985) 224
 [Pisma Zh.\ Eksp.\ Teor.\ Fiz.\  {\bf 42} (1985) 182].
  %%CITATION = JTPLA,42,224;%%

%\cite{Shifman:1985fi}
\bibitem{Shifman:1985fi}
  M.~A.~Shifman, A.~I.~Vainshtein and V.~I.~Zakharov,
  %``Exact Gell-mann-low Function In Supersymmetric Electrodynamics,''
  Phys.\ Lett.\ B {\bf 166} (1986) 334.
  %%CITATION = PHLTA,B166,334;%%

%\cite{Jack:1996vg}
\bibitem{Jack:1996vg}
  I.~Jack, D.~R.~T.~Jones and C.~G.~North,
  %``N=1 supersymmetry and the three loop gauge Beta function,''
  Phys.\ Lett.\ B {\bf 386} (1996) 138
  [hep-ph/9606323].

%\cite{Jack:1996cn}
\bibitem{Jack:1996cn}
  I.~Jack, D.~R.~T.~Jones and C.~G.~North,
  %``Scheme dependence and the NSVZ Beta function,''
  Nucl.\ Phys.\ B {\bf 486} (1997) 479.
  %%CITATION = HEP-PH/9609325;%%

%\cite{Jack:1998uj}
\bibitem{Jack:1998uj}
  I.~Jack, D.~R.~T.~Jones and A.~Pickering,
  %``The Connection between DRED and NSVZ,''
  Phys.\ Lett.\ B {\bf 435} (1998) 61.
  %%CITATION = HEP-PH/9805482;%%



%\cite{Ferrara:1974pu}
\bibitem{Ferrara:1974pu}
  S.~Ferrara and B.~Zumino,
  %``Supergauge Invariant Yang-Mills Theories,''
  Nucl.\ Phys.\ B {\bf 79} (1974) 413.
  %%CITATION = NUPHA,B79,413;%%


%\cite{Jones:1974pg}
\bibitem{Jones:1974pg}
  D.~R.~T.~Jones,
  %``Asymptotic Behavior of Supersymmetric Yang-Mills Theories in the Two Loop Approximation,''
  Nucl.\ Phys.\ B {\bf 87} (1975) 127.
  %%CITATION = NUPHA,B87,127;%%



%\cite{Harlander:2006xq}
\bibitem{Harlander:2006xq}
  R.~V.~Harlander, D.~R.~T.~Jones, P.~Kant, L.~Mihaila and M.~Steinhauser,
  %``Four-loop beta function and mass anomalous dimension in dimensional reduction,''
  JHEP {\bf 0612} (2006) 024
  [hep-ph/0610206].
  %%CITATION = HEP-PH/0610206;%%





%\cite{Shifman:1985tj}
\bibitem{Shifman:1985tj}
  M.~A.~Shifman and A.~I.~Vainshtein,
  %``Operator Product Expansion And Calculation Of The Two Loop Gell-mann-low Function,''
  Sov.\ J.\ Nucl.\ Phys.\  {\bf 44} (1986) 321
   [Yad.\ Fiz.\  {\bf 44} (1986) 498].
  %%CITATION = SJNCA,44,321;%%

%\cite{Mas:2002xh}
\bibitem{Mas:2002xh}
  J.~Mas, M.~Perez-Victoria and C.~Seijas,
  %``The beta function of N=1 SYM in differential renormalization,''
  JHEP {\bf 0203} (2002) 049
  [hep-th/0202082].
  %%CITATION = HEP-TH/0202082;%%

%\cite{Slavnov:1971aw}
\bibitem{Slavnov:1971aw}
  A.~A.~Slavnov,
  %``Invariant regularization of nonlinear chiral theories,''
  Nucl.\ Phys.\ B {\bf 31} (1971) 301.
  %%CITATION = NUPHA,B31,301;%%

%\cite{Slavnov:1972sq}
\bibitem{Slavnov:1972sq}
  A.~A.~Slavnov,
  %``Invariant regularization of gauge theories,''
  Theor.Math.Phys. {\bf 13} (1972) 1064
   [Teor.\ Mat.\ Fiz.\  {\bf 13} (1972) 174].
  %%CITATION = TMFZA,13,174;%%

%\cite{Krivoshchekov:1978xg}
\bibitem{Krivoshchekov:1978xg}
  V.~K.~Krivoshchekov,
  %``Invariant Regularizations for Supersymmetric Gauge Theories,''
  Theor.\ Math.\ Phys.\ {\bf 36} (1978) 745
 [Teor.\ Mat.\ Fiz.\  {\bf 36} (1978) 291].
 %%CITATION = TMFZA,36,291;%%

%\cite{West:1985jx}
\bibitem{West:1985jx}
  P.~C.~West,
  %``Higher Derivative Regulation Of Supersymmetric Theories,''
  Nucl.\ Phys.\ B {\bf 268} (1986) 113.
  %%CITATION = NUPHA,B268,113;%%

%\cite{Soloshenko:2003nc}
\bibitem{Soloshenko:2003nc}
  A.~A.~Soloshenko and K.~V.~Stepanyantz,
  %``Three loop beta function for N=1 supersymmetric electrodynamics, regularized by higher derivatives,''
  Theor.\ Math.\ Phys.\  {\bf 140} (2004) 1264
   [Teor.\ Mat.\ Fiz.\  {\bf 140} (2004) 430].
  %%CITATION = HEP-TH/0304083;%%

%\cite{Pimenov:2009hv}
\bibitem{Pimenov:2009hv}
  A.~B.~Pimenov, E.~S.~Shevtsova and K.~V.~Stepanyantz,
  %``Calculation of two-loop beta-function for general N=1 supersymmetric Yang--Mills theory with the higher covariant derivative regularization,''
  Phys.\ Lett.\ B {\bf 686} (2010) 293.
  %%CITATION = ARXIV:0912.5191;%%

%\cite{Stepanyantz:2011zz}
\bibitem{Stepanyantz:2011zz}
  K.~V.~Stepanyantz,
  %``Quantum corrections in N=1 supersymmetric theories with cubic superpotential, regularized by higher covariant derivatives,''
  Phys.\ Part.\ Nucl.\ Lett.\  {\bf 8} (2011) 321.
  %%CITATION = 00438,8,321;%%

%\cite{Stepanyantz:2011wq}
\bibitem{Stepanyantz:2011wq}
  K.~V.~Stepanyantz,
  %``Factorization of integrals, defining the $\beta$-function, into integrals of total derivatives in N=1 SQED, regularized by higher derivatives,''
  Int.\ J.\ Theor.\ Phys.\  {\bf 51} (2012) 276.
  %%CITATION = ARXIV:1101.2956;%%

%\cite{Smilga:2004zr}
\bibitem{Smilga:2004zr}
  A.~V.~Smilga and A.~Vainshtein,
  %``Background field calculations and nonrenormalization theorems in 4-D supersymmetric gauge theories and their low-dimensional descendants,''
  Nucl.\ Phys.\ B {\bf 704} (2005) 445.
  %%CITATION = HEP-TH/0405142;%%


%\cite{Stepanyantz:2011bz}
\bibitem{Stepanyantz:2011bz}
  K.~V.~Stepanyantz,
  %``Factorization of integrals defining the two-loop $\beta$-function for the general renormalizable N=1 SYM theory, regularized by the higher covariant derivatives, into integrals of double total derivatives,''
  arXiv:1108.1491 [hep-th].
  %%CITATION = ARXIV:1108.1491;%%

%\cite{Stepanyantz:2012zz}
\bibitem{Stepanyantz:2012zz}
  K.~V.~Stepanyantz,
  %``Derivation of the exact NSVZ beta-function in N=1 SQED regularized by higher derivatives by summation of Feynman diagrams,''
  J.\ Phys.\ Conf.\ Ser.\  {\bf 343} (2012) 012115.
  %%CITATION = 00462,343,012115;%%

%\cite{Stepanyantz:2012us}
\bibitem{Stepanyantz:2012us}
  K.~V.~Stepanyantz,
  %``Multiloop calculations in supersymmetric theories with the higher covariant derivative regularization,''
  J.\ Phys.\ Conf.\ Ser.\  {\bf 368} (2012) 012052
  [arXiv:1203.5525 [hep-th]].
  %%CITATION = ARXIV:1203.5525;%%

%\cite{Stepanyantz:2011jy}
\bibitem{Stepanyantz:2011jy}
  K.~V.~Stepanyantz,
  %``Derivation of the exact NSVZ $\beta$-function in N=1 SQED, regularized by higher derivatives, by direct summation of Feynman diagrams,''
  Nucl.\ Phys.\ B {\bf 852} (2011) 71.
  %%CITATION = ARXIV:1102.3772;%%

%\cite{Faddeev:1980be}
\bibitem{Faddeev:1980be}
  L.~D.~Faddeev and A.~A.~Slavnov,
``Gauge Fields. Introduction To Quantum Theory,'' Nauka, Moscow,
1978 and
  Front.\ Phys.\  {\bf 50} (1980) 1
   [Front.\ Phys.\  {\bf 83} (1990) 1].
  %%CITATION = FRPHA,50,1;%%

%\cite{Slavnov:1977zf}
\bibitem{Slavnov:1977zf}
  A.~A.~Slavnov,
  %``The Pauli-Villars Regularization for Nonabelian Gauge Theories,''
  Theor.\ Math.\ Phys.\ {\bf 33} (1977) 977
   [Teor.\ Mat.\ Fiz.\  {\bf 33} (1977) 210].
  %%CITATION = TMFZA,33,210;%%

%\cite{Bogolyubov:1980nc}
\bibitem{Bogolyubov:1980nc}
  N.~N.~Bogolyubov and D.~V.~Shirkov,
  ``Introduction To The Theory Of Quantized Fields,''
Nauka, Moscow, 1984; and Intersci.\ Monogr.\ Phys.\ Astron.\  {\bf
3} (1959) 1.
  %%CITATION = IMTPA,3,1;%%


%\cite{Collins:1984xc}
\bibitem{Collins:1984xc}
  J.~C.~Collins,
  ``Renormalization. An Introduction To Renormalization,
The Renormalization Group, And The Operator Product Expansion,''
  Cambridge, Uk: Univ. Pr. ( 1984) 380p

%\cite{Vladimirov:1975mx}
\bibitem{Vladimirov:1975mx}
  A.~A.~Vladimirov,
  %``Renormalization Group Equations in Different Approaches,''
  Theor.\ Math.\ Phys.\  {\bf 25} (1976) 1170
   [Teor.\ Mat.\ Fiz.\  {\bf 25} (1975) 335].
  %%CITATION = TMPHA,25,1170;%%

%\cite{ArkaniHamed:1997mj}
\bibitem{ArkaniHamed:1997mj}
  N.~Arkani-Hamed and H.~Murayama,
  %``Holomorphy, rescaling anomalies and exact beta functions in supersymmetric gauge theories,''
  JHEP {\bf 0006} (2000) 030
  [hep-th/9707133].
  %%CITATION = HEP-TH/9707133;%%

%\cite{Gorishnii:1990kd}
\bibitem{Gorishnii:1990kd}
  S.~G.~Gorishny, A.~L.~Kataev, S.~A.~Larin and L.~R.~Surguladze,
  %``The Analytical four loop corrections to the QED Beta function in the MS scheme and to the QED psi function: Total reevaluation,''
  Phys.\ Lett.\ B {\bf 256} (1991) 81.
  %%CITATION = PHLTA,B256,81;%%

%\cite{Kataev:2013vua}
\bibitem{Kataev:2013vua}
  A.~L.~Kataev,
  %``Conformal symmetry limit of QED and QCD and the identities between the concrete perturbative contributions to deep-inelastic scattering sum rules,''
  arXiv:1305.4605 [hep-th].
  %%CITATION = ARXIV:1305.4605;%%

%\cite{Kataev:2010tm}
\bibitem{Kataev:2010tm}
  A.~L.~Kataev,
  %``Riemann $\zeta(3)$- terms in perturbative QED series, conformal symmetry and the analogies with structures of multiloop effects in N=4 supersymmetric Yang-Mills theory,''
  Phys.\ Lett.\ B {\bf 691} (2010) 82.
  %%CITATION = ARXIV:1005.2058;%%

%\cite{Soloshenko:2003sx}
\bibitem{Soloshenko:2003sx}
  A.~A.~Soloshenko and K.~V.~Stepanyants,
  %``Two-loop anomalous dimension of N = 1 supersymmetric quantum electrodynamics regularized using higher covariant derivatives,''
  Theor.\ Math.\ Phys.\  {\bf 134} (2003) 377
   [Teor.\ Mat.\ Fiz.\  {\bf 134} (2003) 430].
  %%CITATION = TMPHA,134,377;%%





\end{thebibliography}
\end{document}